\documentclass[usenatbib]{mn2e}
\usepackage{txfonts}
\usepackage{epsfig}
\usepackage{graphicx}
\usepackage{natbib}
\usepackage{subfigure}
\usepackage{ctable}
\usepackage{appendix}
\title[Statistical searches for low-frequency solar oscillations]
{A comparison of frequentist and Bayesian inference: Searching for
low-frequency p modes and g modes in Sun-as-a-star data}

\author[A.-M. Broomhall et al.]{A.-M.
Broomhall$^1$\thanks{amb@bison.ph.bham.ac.uk}, W.~J. Chaplin$^1$, Y.
Elsworth$^1$, T. Appourchaux$^2$, R. New$^3$\\ $^1$School of Physics
and Astronomy,
University of Birmingham, Edgbaston, Birmingham B15 2TT\\
$^2$Institut d'Astrophysique Spatiale, CNRS-Universit\'{e} Paris XI
UMR 8617, 91405 Orsay Cedex, France\\
$^3$Faculty of Arts, Computing, Engineering and Sciences, Sheffield
Hallam University, Sheffield S1 1WB}
\begin{document}

\maketitle
\begin{abstract}
We describe and use two different statistical approaches to try and
detect low-frequency solar oscillations in Sun-as-a-star data: a
frequentist approach and a Bayesian approach. We have used
frequentist statistics to search contemporaneous Sun-as-a-star data
for coincident, statistically-prominent features. However, we find
that this approach leads to numerous false detections. We have also
used Bayesian statistics to search for evidence of low-frequency p
modes and g~modes in Sun-as-a-star data. We describe how Bayesian
statistics can be used to search near-contemporaneous data for
coincident prominent features. Near-contemporaneous data were used
to circumvent the difficulties in deriving probabilities that occur
when common noise is present in the data. We find that the Bayesian
approach, which is reliant on the assumptions made when determining
the posterior probability, leads to significantly fewer false
detections and those that are observed can be discredited using a
priori knowledge. Therefore, we have more confidence in the mode
candidates found with Bayesian statistics.
\end{abstract}


\section{Introduction}

Low-frequency p~modes are expected to have very long lifetimes and
so, if detected, their frequencies can be determined both accurately
and precisely. This, in turn, makes low-frequency p~modes important
for constraining~models of the structure of the solar interior. Only
p~modes with low degrees (low $l$) travel through regions close to
the solar centre and so observations of low-$l$ p~modes are
important for inferring the structure of the solar core. Mode
amplitudes decrease with frequency and this, coupled with the
increasing level of noise at lower frequencies, means that below
$1500\,\rm\mu Hz$ oscillation signatures are difficult to detect. No
independently confirmed detections of the core-penetrating, low-$l$
p~modes have been made below $\sim970\,\rm\mu Hz$
\citep[e.g.][]{Garcia2001, Chaplin2002, Broomhall2007}.

The large sound speed in the solar core means that the effect of
core conditions on low-degree p~modes is relatively small. On the
other hand, solar gravity (g) modes, which have frequencies below
$450\,\rm\mu Hz$, are confined to the radiative interior and core,
making them an extremely sensitive probe of the deep solar interior
\citep[see e.g.][]{Garcia2008, Mathur2008}. Since g~modes are
evanescent outside the radiative interior their signatures are
significantly attenuated by the time they reach the photosphere,
where\vspace{0.5cm} helioseismic observations are made. It is,
therefore, a major observational challenge to observe these modes.
\citet{Appourchaux2010} is an up-to-date review of the current
status of the search for g~modes. \citet{Garcia2007} claim to have
found evidence of the signatures of g~modes in data observed by the
Global Oscillations at Low Frequencies (GOLF) instrument, onboard
\emph{Solar and Heliospheric Observatory (SOHO)}. \citet{Garcia2008}
found similar results when they examined 20 per cent more GOLF data
than used by \citet{Garcia2007}. However, these results await
independent confirmation and, to date, there have been no detections
of individual g~modes.

Some solar oscillations are thought to show a mixed character
\citep*[e.g.][]{Provost2000}. They have a sensitivity to the Sun's
internal structure that looks like a g~mode, and so have excellent
diagnostic potential, but they respond to surface conditions in a
manner that is similar to p~modes. Therefore, theoretically, mixed
modes should be easier to detect than g~modes, however, no mixed
modes have yet been definitively identified. \citet{Garcia2001} and
\citet{Gabriel2002} tentatively identify one member of the $n=1$,
$l=1$ p~mode, which is thought to be mixed. However, they are unable
to determine which azimuthal component is detected.

Any signal from a low-frequency p~mode (below $1500\,\rm\mu Hz$) or
g~mode is expected to be small compared to the surrounding
background noise and so the following question arises: How does an
observer decide whether a detection is real or not? This is where
statistical testing becomes important as it allows an observer to
quantify the significance of any potential detection. To date the
majority of statistical-based searches for low-degree p~modes and
g~modes have taken a frequentist approach, i.e. it is common to
determine the probability of observing a certain feature in a
frequency-power spectrum based on the hypothesis that the spectrum
contains pure noise. This is the H0 hypothesis. However, recent work
by \citet{Appourchaux2008} implies that the `$P$ values', which are
quoted to assess the significance of a detection in a frequentist
approach, may be too optimistic. It is, therefore, useful to also
determine the probability of observing a feature in a
frequency-power spectrum based on the alternative H1 hypothesis. The
H1 hypothesis assumes that there is a signal in the data. However,
to calculate this probability is it necessary to make certain
\emph{a priori} assumptions about the signal and so a Bayesian
approach is appropriate. \citeauthor{Appourchaux2008} proposes that
Bayesian statistics provides a more rigourous method of assessing
the significance of any mode candidates than a frequentist approach.
When using Bayesian statistics it is possible to use our knowledge
of the structure of the Sun to put more constraints on the search
for solar oscillations than when using a frequentist approach. In
this paper we use both frequentist and Bayesian statistics to search
for low-frequency p~modes, g~modes and mixed modes.

We have taken care to minimize the number of false detections. There
are two ways in which false detections can be made. A type~I error
occurs when the null hypothesis is wrongly rejected. Here, that
would mean wrongly rejecting the hypothesis that a prominent
structure is due to noise. In other words a type~I error would occur
if we flagged a structure as a mode candidate when it is noise. A
type~II error occurs when the null hypothesis is not rejected when,
in fact, it is false. In this paper that would mean accepting the
hypothesis that a feature is due to noise when it is not. We regard
type~I errors as more serious than type~II errors. Therefore we have
decided to err on the side of caution and set very stringent
requirements that need to be passed before a structure is considered
as a mode candidate. For example, we require that any mode
candidates be close to the mode frequencies predicted by solar
models. While this restriction may increase the number of type~II
errors it is also likely to restrict the number of type~I errors,
which we consider to be more important.

The structure of this paper is as follows: we begin, in
Section~\ref{section[comparison]}, with a comparison of the
frequentist and Bayesian approaches. We then give a more detailed
description of how frequentist statistics were used to search
various data sets (Section \ref{section[frequentist method]}). We
have compared contemporaneous data for coincident features that are
prominent when compared to the background noise. When determining
the probability of observing such features in contemporaneous data,
it is necessary to take account of any common noise that is present
in the data due to, for example, the solar granulation. In order to
do this \citet{Broomhall2007} derived a joint probability based on
the assumption that the background noise has a Gaussian
distribution, which is true in frequency-amplitude spectra.
Therefore, when using the frequentist approach to search for
low-frequency solar oscillations we have examined
frequency-\emph{amplitude} spectra.

In Section \ref{section[Bayes intro]} we give a brief introduction
to Bayesian statistics and describe how they can be used to search
for low-frequency oscillations in solar data. Bayesian statistics
can be used to determine the `posterior probability' that a
hypothesis, such as H0, is true given a set of observed data. The
derivation of the posterior probability is based on the assumption
that the background noise has a $\chi^2$, 2 degrees of freedom
(d.o.f.) distribution, which is true in frequency-power spectra.
Therefore, when using a Bayesian approach to search for
low-frequency solar oscillations we have examined
frequency-\emph{power} spectra. In Section \ref{section[prior]} we
discuss one component of the Bayesian calculations: the prior
probability. In particular we determine whether this probability can
make use of a priori knowledge to guide searches for solar
oscillations. To date Bayesian statistics have only been used to
search individual data sets for evidence of oscillations. We have
used a Bayesian approach to derive the posterior probability that a
coincident prominent feature will be detected in data observed by
different instruments (Section \ref{section[Bayes stats]}).

We have searched three sets of data for evidence of low-frequency
oscillations using both frequentist and Bayesian approaches. The
data were observed by the Birmingham Solar Oscillations Network
(BiSON), and the GOLF and Michelson Doppler Imager (MDI) instruments
onboard the \emph{SOHO} spacecraft. The results of searching the
data are described in Section \ref{section[results]}. A comparison
of the results of the two different approaches is made in Section
\ref{section[discussion]}.

\section{A general comparison between frequentist and Bayesian
statistics}\label{section[comparison]} When using a frequentist
approach to search for low-frequency solar oscillations the
probability of observing, by chance, a prominent feature in a range
of $N$ frequency bins in a frequency-amplitude/ power spectrum is
often determined. The determined probability is dependent on the
value of $N$, which must be chosen carefully and the choice of $N$
is discussed in more detail in Section~\ref{section[frequentist
method]}.

When adopting a frequentist approach to statistics it is common to
set, a priori, an arbitrary probability detection threshold level,
for example, at 1 per cent. If the probability of observing a
prominent feature is less than this threshold probability the
feature is considered as a mode candidate. However, the meaning of
the threshold probability is commonly misunderstood. Take, for
example, the frequentist approach that was used here. We determined
the probability of observing a prominent spike in a
frequency-amplitude spectrum by chance, if the H0 hypothesis is
true. The H0 hypothesis assumes that the data being examined contain
only noise. However, the determined probability, or $P$ value, is
commonly misinterpreted as the probability that the H0 hypothesis
itself is true. In fact, we can say nothing about the validity of
the hypothesis.

The problem occurs because of the non-repeatability of the
experiment: we only have one Sun to observe. A similar problem has
been identified in medical statistics \citep[e.g.][]{Berger1987,
Sellke2001}, where it has been shown that the probability that the
H0 hypothesis is correct is significantly larger than the $P$ value
quoted in a frequentist approach. Consequently, great care must be
taken when drawing conclusions from frequentist statistics. It must
be made clear the frequentist threshold probability gives the
chances of the data being observed if the underlying hypothesis is
true and gives no information on the validity of the hypothesis
itself.

To determine the probability that a given hypothesis is true we must
turn to Bayesian statistics. Bayesian statistics allow us to use~a
priori information in a statistically rigourous manner to compare
the chances that, given the observed data, different hypotheses are
true. For example, here we have used Bayesian statistics to
determine which of the H0 and H1 hypotheses are most likely, where
the H1 hypothesis assumes there is a signal in the data. It is
important to point out that we are still not determining the
probability that a detection is a solar mode of oscillation as the
H1 hypothesis does not specify the source of the signal. Any
observed signal could conceivably be some sort of structured noise.
However, the Bayesian approach still gives probabilities that are
less open to misinterpretation than the frequentist approach.
Furthermore, significance estimates are much more conservative than
those obtained with frequentist methods and so should lead to fewer
false detections.

\section{Searching contemporaneous data using frequentist statistical
techniques}\label{section[frequentist method]} The signal from solar
oscillations will be common to contemporaneous data observed by
different instruments. It is, therefore, pertinent to search
frequency-amplitude spectra, constructed from contemporaneous data,
for statistically significant, prominent concentrations of power,
which lie significantly above the local noise background and that
are coincident in frequency in different data sets. In Section
\ref{section[results]} we show the results obtained when
contemporaneous BiSON, GOLF and MDI data were searched.

It is possible to determine the probability of observing, by chance,
prominent features that are coincident in frequency in two
frequency-amplitude spectra. Proper account must be taken of any
noise that is common to the two sets of data since the presence of
common noise increases the probability of observing a coincident
prominent noise feature. When BiSON data are compared with GOLF or
MDI data any common noise is solar in origin and, more specifically,
is due to solar granulation. When GOLF and MDI data are compared
some of the common noise may also be instrumental as both are
onboard the SOHO spacecraft. The presence of common noise means that
the probability of observing coincident prominent features by chance
is non-trivial to derive. A derivation of the joint probability,
$p$, of the occurrence of coincident prominent features in two
frequency-amplitude spectra is given in \citet{Broomhall2007}. In
order to make this derivation it was assumed that the background is
slowly varying in frequency with no sharp features and that the real
and imaginary noise in the frequency-amplitude spectrum has a
Gaussian distribution, as is observed.

The probability of observing, by chance, a coincident, prominent
feature at least once over a region of $N$ bins in two
frequency-amplitude spectra is
%%%%%%%%%%%%%%%%%%%%%%%%%%%%%%%%%%%%%%%%%%%%%%%%%%%%%%%%%%%%%%%
\begin{equation}\label{equation[prob over N bins]}
    P=1-(1-p)^N,
\end{equation}
%%%%%%%%%%%%%%%%%%%%%%%%%%%%%%%%%%%%%%%%%%%%%%%%%%%%%%%%%%%%%%%
where again we recall that $p$ is the joint probability for finding
a coincident feature in a single bin. A low value of $P$ indicates
that a feature is statistically prominent compared to the background
noise and, as such, is worth consideration as a `mode candidate'. It
is important to remember that this is fundamentally different from
determining the probability that a detection is a mode.

When determining the significance of any observed prominent features
the value of $N$ in equation (\ref{equation[prob over N bins]}) must
be fixed. As $N$ increases, the probability of detecting a prominent
feature by chance in a single set of $N$ bins increases and
therefore the size of $N$ must be suitably capped. However, as $N$
decreases the total number of strips containing $N$ bins in the
frequency range of the frequency-amplitude spectrum that is searched
increases. Consequently, the probability that at least one of these
$N$ bin ranges contains a false detection also increases and so $N$
should not be too small. When deciding on the optimum value of $N$ a
balance between these two requirements must be sought. In the
results that follow $N$ was taken to be the number of bins in
$100\,\rm\mu Hz$.

Various statistical tests are described in \citet{Chaplin2002} that
determine the probability of observing prominent structures in a
single frequency-power spectrum. These tests can be easily adapted
so they can be used to compare two frequency-amplitude spectra if
the probability of observing a single prominent spike at the same
frequency in each of the frequency-amplitude spectra can be
calculated \citep{Broomhall2007}. These tests were originally
designed to detect p~modes, however, the underlying theory is still
applicable when searching for g~modes. An outline of the statistical
tests that were applied here will now be given.

We define a spike as a single bin with a prominent amplitude in a
frequency-amplitude spectrum. A `pair of spikes' is then a prominent
spike that occurs in the same frequency bin in each~of two
frequency-amplitude spectra. The simplest test that was used to
compare the BiSON, GOLF and MDI data searched for individual
prominent pairs of spikes. This test makes no a priori assumptions
regarding the properties of the modes. Additional tests have also
been formulated which do take advantage of known mode properties
that are observed in frequency-power/amplitude spectra.

One such test searches for a cluster of prominent spikes. For
example, we have searched for three coincident prominent spikes over
a range of six bins. This test is designed to take advantage of the
fact that solar oscillations are damped and the power of a mode may
be spread across more than one frequency bin. The observed width of
a mode is dependent on the mode's lifetime and the length of the
respective data set. Although the lifetimes of low-frequency p~modes
are uncertain, evidence from higher-frequency p~modes implies that
in a time series of length $8.5\,\rm yrs$, which is the length of
the data series analysed here, even very low-frequency p~modes could
have resolved widths \citep{Broomhall2008}. The lifetimes of g~modes
are less predictable and so it is uncertain whether or not the width
of g~modes will be resolved. Even if a mode's lifetime is greater
than the length of the time series it is still possible for its
power to be spread across more than one bin for the following
reason. If the mode signal is not commensurate with the window
function of the observations, its maximum amplitude will be
diminished and the majority of its amplitude will be divided between
two consecutive frequency bins. With this in mind, a statistical
test was also derived that searched for two consecutive prominent
pairs of spikes.

Another statistical test that was used to search the BiSON, GOLF and
MDI data looked for prominent structures whose patterns mimic those
expected from the rotational splitting. In Sun-as-a-star data only
mode components where $l+m$ is even are visible and so the
statistical test searched for components whose azimuthal orders,
$m$, were separated by 2. This test assumed a separation in
frequency between adjacent $m$ components of $0.4\pm0.1\,\rm \mu
Hz$. The error of $\pm0.1\,\rm \mu Hz$ is included to allow for the
influence of magnetic fields, which shift the observed frequencies
of the modes. \citet{Garcia2004} found that the splittings of
p~modes can vary between 0.38 and $0.45\,\rm\mu Hz$, both of which
are within the range of splittings allowed here i.e. $0.4\pm0.1\,\rm
\mu Hz$.

We use a splitting of $0.4\pm0.1\,\rm \mu Hz$ to maintain
consistency with previous works \citep{Chaplin2002, Broomhall2007}.
However, it should be noted that since the rotation in the core is
not well known this value may not represent the synodic splitting of
g~modes. For example, \citet{Garcia2007} suggest that the core may
be rotating between 3 and 5 times faster than the radiative zone. We
note here that one could question the how appropriate it is to use
this test with a frequentist approach since we are using a priori
knowledge. Therefore, although the results of the multiplet test are
included here for completeness and consistency with previous works
\citep{Chaplin2002, Broomhall2007} we regard the results with
suspicion.

More detailed descriptions of the statistical tests can be found in
\citet{Chaplin2002} and \citet{Broomhall2007}. We now describe how
Bayesian statistics can be used to assess the significance of any
mode candidates.

\section{Using Bayesian statistics to search for solar
oscillations}\label{section[Bayes intro]} \citet{Appourchaux2008}
describes how Bayesian statistics can be used to determine the
frequencies of modes in solar data and \citet{Appourchaux2009} uses
Bayesian statistics to search asteroseismic data observed by the
COnvection ROtation and planetary Transits (CoRoT) spacecraft for
evidence of oscillations. We now summarize the methods used by
\citet{Appourchaux2009} and adapt them so that they can be used to
search contemporaneous solar data.

Bayes theorem can be used to determine the `posterior probability',
$p(\textrm{H}0|x)$, that the H0 hypothesis is true given the
observed data, $x$. \citet{Appourchaux2009} shows that the posterior
probability is given by
%%%%%%%%%%%%%%%%%%%%%%%%%%%%%%%%%%%%%%%%%%%%%%%%%%%%%%%%%%%%%%%%%%%%%%%%
\begin{equation}\label{equation[Bayes reduced]}
p(\textrm{H}0|x)=\left[1+\frac{(1-p_0)}{p_0}\frac{p(x|\textrm{H}1)}{p(x|\textrm{
H}0)}\right]^{-1},
\end{equation}
%%%%%%%%%%%%%%%%%%%%%%%%%%%%%%%%%%%%%%%%%%%%%%%%%%%%%%%%%%%%%%%%%%%%%%%%%
where $p_0$ is the prior probability, which is defined as the
subjective probability, before any observations have been made, that
H0 is true. The H0 hypothesis assumes that the data contains only
noise. Therefore, in equation (\ref{equation[Bayes reduced]}),
$p(x|\textrm{H}0)$ is the probability that a spike with a power $x$
is observed in a frequency-power spectrum that contains only noise.
Similarly, $p(x|\textrm{H}1)$ is the probability that a spike with a
power, $x$, is observed if the $\textrm{H}1$ hypothesis is true.
Here, the $\textrm{H}1$ hypothesis assumes that there is a signal in
the data with a maximum power spectral density per unit bin, $H$. We
are assuming that H0 and H1 cover all possibilities. In other words
there is no third hypothesis, H2 say, that is true.
\citet{Appourchaux2009} derives the expressions for
$p(x|\textrm{H}0)$ and $p(x|\textrm{H}1)$ that can be used to search
frequency-power spectra. We now quote these expressions for
reference.

We define a spike as the power spectral density contained in a
single bin of the frequency-power spectrum. Frequency-power spectra
have a $\chi^2$, 2 d.o.f. distribution and so the probability of
observing a spike with a power spectral density of $x$ if the data
contain noise only is given by
\begin{equation}\label{equation[p(x|h0)]}
    p(x|\textrm{H}0)=\textrm{e}^{-x},
\end{equation}
where we have taken the mean power spectral density of the spectrum
to be 1. For the alternative H1 hypothesis we assume that the spike
contains the signal from a mode which has been stochastically
excited, as is the case for solar oscillations. However, as we do
not know the power of the mode we assume, a priori, that the maximum
power spectral density per unit bin of the mode in a frequency-power
spectrum is distributed uniformly between $0$ and $H$. The
probability of observing a spike with a power spectral density, $x$,
is then given by
\begin{equation}\label{equation[p(x|h1)]}
p(x|\textrm{H}1)=\frac{1}{H}\int_0^H\frac{1}{1+H'}\textrm{e}^{-x/(1+H')}\textrm{
d}H'.
\end{equation}
Equations (\ref{equation[p(x|h0)]}) and (\ref{equation[p(x|h1)]})
can be substituted into equation (\ref{equation[Bayes reduced]}) to
give the posterior probability.

\subsection{Range of heights}\label{section[amplitude range]}
In order to calculate $p(x|\textrm{H}1)$ we have to make assumptions
about the maximum power spectral density per unit bin of a mode
profile, $H$, with a frequency, $\nu$. Fig. \ref{figure[post prob vs
a]} shows the variation with $H$ of the posterior probability, which
is related to $p(x|\textrm{H}1)$ by equation (\ref{equation[Bayes
reduced]}), for power spectral densities, $x$. The choice of $H$ is
most influential at low $H$ and high $x$ and so it is important to
choose the value of $H$ carefully to enable the significance of any
prominent feature to be accurately assessed. As we are searching for
oscillations that have not yet been observed we need to use models
to estimate the value of $H$.

Notice that for each $x$ plotted in Fig. \ref{figure[post prob vs
a]} a minimum posterior probability is observed. Such minima can be
understood as a manifestation of the fact that the prior on $H$
being large is contradicted by the low power spectral density level
reached by the peak \citep{Appourchaux2009}. In other words, if the
mode power spectral densities were truly so high the observed peak
would also be prominent.

\begin{figure}
  \centering
  \includegraphics[width=0.4\textwidth, clip]{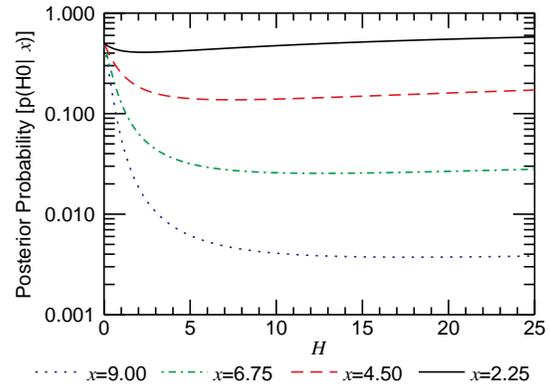}\\
  \caption{The variation of the posterior probability with the
  maximum power, $H$, for different threshold powers, $x$.}
  \label{figure[post prob vs a]}
\end{figure}

Before we discuss the values used to constrain $H$ it is important
to define exactly what we mean by the maximum power spectral density
per unit bin of a mode. In this paper we define the maximum power
spectral density as the peak height of a mode's profile in a
frequency-power spectrum. Solar oscillations are damped and, if the
lifetime of a mode is significantly shorter than the length of the
observations, the mode will appear in a frequency-power spectrum in
the form of a Lorentzian. The width of the Lorentzian, $\Delta$, is
related to a mode's lifetime, $\tau$, by the following equation:
%%%%%%%%%%%%%%%%%%%%%%%%%%%%%%%%%%%%%%%%%%%%%%%%%%%%%%%%%%%%%%%%%%%%%%
\begin{equation}\label{equation[width-lifetime]}
\Delta=\frac{1}{\pi\tau}.
\end{equation}
%%%%%%%%%%%%%%%%%%%%%%%%%%%%%%%%%%%%%%%%%%%%%%%%%%%%%%%%%%%%%%%%%%%%%%%
The maximum power spectral density per unit bin (or height) can be
described by the following equation \citep{Fletcher2006}:
%%%%%%%%%%%%%%%%%%%%%%%%%%%%%%%%%%%%%%%%%%%%%%%%%%%%%%%%%%%%%%%%%%%%%%%%%
\begin{equation}\label{equation[intermediate H]}
H=\frac{2V^2}{\pi T\Delta+2}.
\end{equation}
%%%%%%%%%%%%%%%%%%%%%%%%%%%%%%%%%%%%%%%%%%%%%%%%%%%%%%%%%%%%%%%%%%%%%%%%%%
where $V^2$ is the total power of the mode, $T$ is the length of the
time series and $\Delta$ is in units of frequency. Consequently,
$T\Delta$ gives the width of the Lorentzian in units of bins.

For p~modes the value of $H$ was estimated in two different ways,
each of which will now be described. \citet{Broomhall2008}
extrapolated to low frequencies the total mode powers, $V^2$, and
widths, $\Delta$, of well-defined, higher-frequency modes, whose
profiles in a frequency-power spectrum can be fitted accurately. The
total power, $V^2$, was extrapolated in preference to $H$ as it is
not dependent on $\Delta$. \citeauthor{Broomhall2008} did this in
two different ways and here we have used the method that produced
the largest $H$ values.

Even so, it is possible that this method under-estimates the power
of the signal from a mode. For example, \citet{Broomhall2008} found
that the $l=0$, $n=6$ mode at $\sim972\,\rm\mu Hz$ is excited to a
larger power than is expected from the extrapolations. In fact, the
simulations performed by \citeauthor{Broomhall2008} indicated that,
if the actual power was equal to the value given by the
extrapolation, the probability that the $l=0$, $n=6$ mode would be
excited to the observed power is approximately 0.01. Furthermore,
because solar oscillations are excited stochastically, the total
power of a mode, $V^2$, varies with time. If the value of the power
given by the extrapolation is regarded as a mean, the observed power
of a mode is equally likely to be greater than this value as it is
to be lower. Consequently, by integrating over the range 0 to $H$ we
could be underestimating the true height of the mode. More realistic
probabilities may be obtained if we overestimate the range of
heights over which the integration in equation
(\ref{equation[p(x|h1)]}) is made. Therefore, we have also performed
a search where the maximum power spectral density per unit bin, $H$,
was assumed to be constant at all frequencies and to have the height
predicted by \citet{Broomhall2008} for a mode with a frequency of
$1500\,\rm\mu Hz$. A larger value of $H$ should allow signals to be
detected more easily as when $H$ tends to 0, $p(x|\textrm{H}1)$
tends to $p(x|\textrm{H}0)$ and so it is very difficult to determine
which hypothesis is most likely to be correct (see Fig.~
\ref{figure[post prob vs a]}).

As for low-frequency p~modes we are reliant on models to predict the
powers of g~modes. Two groups have produced predictions of g-mode
powers. The first group uses the Cambridge stochastic excitation
model, as applicable to g~modes \citep{Houdek1999}. The second group
\citep{Belkacem2009} uses a similar model with a different
temporal-correlation between the modes and the turbulent eddies:
\citeauthor{Belkacem2009} take the eddy-time correlation function to
be a Lorentzian, while \citeauthor{Houdek1999} assume that the
eddy-time correlation is a Gaussian. The authors of both papers feel
that their respective models are only applicable to limited ranges
in frequency and these ranges do not coincide. However, the work of
\citeauthor{Belkacem2009} predicts significantly larger g-mode
powers than those predicted by \citeauthor{Houdek1999}. Therefore,
we have taken the upper limit for $H$ to be the maximum power
predicted by the work of \citet{Belkacem2009} as the larger the mode
powers the more likely we are to be able to detect any oscillations.
We have taken the total power, $V^2$, to be the constant value of
$9\,\rm mm^2\, s^{-2}$ at all frequencies in the range
$(50-350)\,\rm\mu Hz$. We have then assumed that the lifetime of the
g~modes is significantly longer than the length of the observations
and so all of the modes power will be confined to a single bin.
Therefore the height of the mode, $H$, is equal to the total power,
$V^2$ i.e. $H=9\,\rm mm^2\, s^{-2}\,bin^{-1}$. \citet{Belkacem2009}
note that, given the uncertainties in their model, this maximum
power could, potentially, be a factor of four larger. Therefore, we
have also determined the posterior probability given the observed
prominent features when the maximum height, $H$, was $36\,\rm
mm^2\,s^{-2}\,bin^{-1}$.

\subsection{Searching a range of frequencies}
The statistics described above can be used to search individual bins
of a frequency-power spectrum. However, there are $\sim400,000$ bins
in the total range of frequencies searched here $(50-1500\,\rm\mu
Hz)$. Consequently the chances of making a false detection are
significant when searching each bin individually. We therefore
consider a range of frequencies containing $M$ bins. In the manner
described by \citet{Sturrock2008} we define the following null and
alternative hypotheses:
\newline
\newline
\noindent H00: The principal peak in $M$ bins is due to
noise,\newline \noindent H11: The principal peak is due to an
oscillatory signal with a frequency, $\nu$.\newline
\newline
\noindent Then
\begin{equation}\label{equation[p(h00)]}
P(x|\textrm{H}00)=1-[1-P(x|\textrm{H}0)]^M\approx MP(x|\textrm{H}0),
\end{equation}
when $P(x|\textrm{H}0)$ is small. We assume only one frequency bin
out of the $M$ of interest contains a signal and so
\begin{equation}\label{equation[p(h11)]}
P(x|\textrm{H}11)=P(x|\textrm{H}1).
\end{equation}
Therefore,
\begin{eqnarray}\label{equation[p(h00|x)]}
p(\textrm{H}00|x)&=&\left[1+\frac{(1-p_0)}{p_0}\frac{p(x|\textrm{H}11)}{
p(x|\textrm{H}00)}\right]^
{-1}\nonumber\\
&=&\left[1+\frac{(1-p_0)}{p_0}\frac{p(x|\textrm{H}1)}{Mp(x|\textrm{H}0)}\right]^
{-1}.
\end{eqnarray}
To ensure that only one mode component is present in the $M$ bins we
have taken $M$ to be the number of bins in $0.4\,\rm\mu Hz=106$, as
this is the expected splitting between adjacent $m$ components. Note
that we have erred on the side of caution when setting this value of
$M$ as in Sun-as-a-star data only $(l+m)$-even components can be
detected. Therefore, the average splitting between adjacent visible
components is closer to $0.8\,\mu\rm Hz$. We have assumed that the
components of g~modes are also rotationally split by $0.4\,\rm \mu
Hz$. It should be noted that the core rotation rate is very
uncertain and so this assumption may not be valid. For example,
\citet{Garcia2007} suggest that the core might be rotating between
three and five times faster than the radiative zone. However, even
if this is the case there will still be only one component present
in the $M=106$ bins under consideration. One possible extension to
this work would be to test the data taking into account a faster
core rotation rate.

\subsection{Searching smoothed spectra using Bayesian
techniques}\label{section[smoothed bayes]} The lifetimes of p~modes
are expected to be short enough for their profiles to have a
resolved width in the frequency-power spectrum examined here. Until
now we have assumed that g~modes have very long lifetimes but this
is not necessarily correct. \citet{Garcia2007} have found a very
interesting peak in a periodogram of a periodogram of GOLF data.
They claim that this peak is due to the superposition of g~modes
that lie in the asymptotic range where the separation in period is
approximately constant ($\sim24\,\rm min$). The peak is positioned
close to the period predicted by standard solar models.
\citeauthor{Garcia2007} use artificial data to show that, if this
peak is evidence of g~modes, the modes might have widths in a
frequency-power spectrum that are commensurate with damping times of
several months. However, this is not the only explanation for the
observed mode widths. The observed results could also be due to an
internal magnetic splitting, changes with time of the cavities in
which the oscillations were trapped or a noise effect in the GOLF
data. If the damping times are of the order of months and given that
the data searched here are $3071\,\rm d$ in length, the width of
these modes will be resolved in frequency-power spectra.

\citet{Appourchaux2009} describe how Bayesian statistics can be used
to search for modes with resolved widths. This is done by summing
the frequency-power spectrum over a given number of bins. Smoothing
a spectrum alters the underlying statistics of the noise and so it
is necessary to derive new equations for $p(x|\textrm{H}0)$ and
$p(x|\textrm{H}1)$. If we smooth a frequency-power spectrum over $R$
bins the underlying statistics of the noise becomes $\chi^2$ with
$2R$ d.o.f.. We have used the equations for $p(x|\textrm{H}0)$ and
$p(x|\textrm{H}1)$, which are described in Section 3.2 of
\citet{Appourchaux2009}, to search solar data for modes with
lifetimes that are shorter than the length of the time series. We
now give the equations for reference.

The probability that, in a spectrum that has been smoothed over $R$
bins, a spike with a power spectral density of $x$ is observed if
the $\textrm{H}0$ hypothesis is true is given by
\begin{equation}\label{equation[smoothed p(x|h0)]}
   p(x|\textrm{H}0)=\frac{x^{R-1}\textrm{e}^{-x}}{\Gamma(R)},
\end{equation}
where $\Gamma(R)$ is the gamma function and we have assumed that the
mean of the unsmoothed frequency-power spectrum is unity. For the
alternative hypothesis we again assume that the spike contains the
signal from a stochastically excited mode. However, when dealing
with smoothed spectra, instead of making assumptions about $H$ we
must make assumptions about the total mode power, $V^2$, and the
mode linewidth, $\Delta$. Here we assumed, a priori, that the mode
amplitude is distributed uniformly between 0 and $V$ and that the
mode linewidth is distributed uniformly between 0 and $\Delta$. Then
the probability of observing a spike with a power spectral density
$x$ in a smoothed spectrum, if the $\textrm{H}1$ hypothesis is true,
is given by
\begin{equation}\label{equation[smoothed p(x|h1)]}
    p(x|\textrm{H}1)=\frac{1}{V\Delta}\int^V_0\int^\Delta_0\frac{\lambda^\nu}
    {\Gamma(\nu)}x^{\nu-1}\textrm{e}^{-\lambda
    x}\textrm{d}V'\textrm{d}\Delta',
\end{equation}
where $\lambda$ and $\nu$ are given in \citet{Appourchaux2004} and
are related to the line profile of the mode, i.e. they are related
to $V$ and $\Delta$. Equations (\ref{equation[smoothed p(x|h0)]})
and (\ref{equation[smoothed p(x|h1)]}) can be substituted into
equation (\ref{equation[p(h00|x)]}) to give the posterior
probability.

When searching for p~modes, we have used the values for $V^2$ and
$\Delta$ extrapolated by \citet{Broomhall2008} as the upper limits
on the mode powers and lifetimes. When searching for g~modes we took
$V^2$ to be both $9\,\rm mm^2\,s^2$ and $36\,\rm mm^2\,s^2$ (see
Section \ref{section[amplitude range]} for details). As the
lifetimes of the g~modes are not known we used a range of different
values: 1, 2, 3, and 4 months. Equation (\ref{equation[p(h00|x)]})
was then used to determine the posterior probability. The posterior
probability, given by Equation (\ref{equation[p(h00|x)]}), is also
dependent on the prior probability, $p_0$, and we now discuss the
value given to $p_0$.

\section{What value should the prior probability
take?}\label{section[prior]}

One important question when applying Bayesian statistics is what
value should the prior probability, $p_0$, take? We define $p_0$ as
the subjective probability that the $\textrm{H}0$ hypothesis is
correct, as assigned before the data are examined. If $p_0$ is set
at a very high (or very low) value any statistical tests are likely
to just confirm the initial belief that $\textrm{H}0$ (or
$\textrm{H}1$) is correct. Therefore, it is common to set $p_0=0.5$
so as to avoid prejudicing one hypothesis over the other. However,
we know that low-frequency p~modes have narrow widths, which cover
at most a few bins only. In which case, would we expect the
probability that $\textrm{H}1$ and $\textrm{H}0$ are true to be the
same at all frequencies? Is it possible to use our knowledge of the
solar structure to tell us which hypothesis is more likely to be
true at a given frequency? Can we also use this knowledge to then
systematically guide the regions in frequency we wish to search?

\subsection{A non-uniform prior that can be used to search for p~modes}

\begin{figure}
  \centering
  \includegraphics[width=\columnwidth]{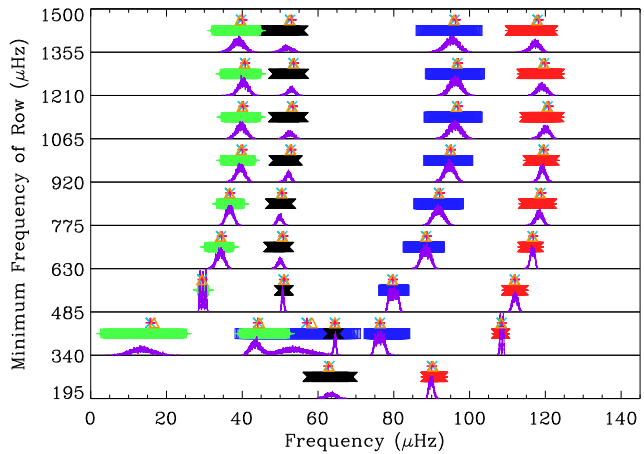}\\
  \caption{An echelle diagram showing the frequencies predicted by
  different solar models. At the centre of each row, the black symbols represent
the
  frequencies of the $l=0$ modes, the red symbols represent the
  $l=1$ modes, the green symbols represent the $l=2$ modes and the
  blue symbols represent the $l=3$ modes. All components with
  $l+m$ even have been plotted assuming a symmetric synodic
  splitting of $0.4\,\rm\mu Hz$. The purple curve shows the frequency
  distribution of the models. Towards the top of each row the $m=0$
  components of the Saclay seismic model frequencies \citep[][blue
crosses]{Turck-Chieze2001}, the M1 model \citep[][orange
triangles]{Zaatri2007}
  and Model S frequencies \citep[][pink plus signs]{Christensen1991} are shown
for
  comparison purposes.}\label{figure[model echelle]}
\end{figure}

We have looked at the p-mode frequencies predicted by 5000 solar
models (see \citealt*{Bahcall2006}; \citealp{Chaplin2007}, for
details), whose input parameters, such as composition and age, have
been altered within their error bars. The frequencies of the modes
predicted by the models have been plotted on an echelle diagram in
Fig. \ref{figure[model echelle]}. Different colours have been used
to represent the different $l$ and the predicted frequencies of all
of the visible components (when $l+m$ is even) are plotted. We have
taken the synodic splitting to be $0.4\,\rm\mu Hz$ and symmetric,
which is a reasonable assumption for the frequency range of interest
here \citep[$<1500\,\rm\mu Hz$, ][]{Chaplin2002}. Although the
rotation rate of the solar core is uncertain this assumption remains
valid as the p~mode rotational splitting is relatively insensitive
to the rotation speed of the core \citep[see e.g.][]{Garcia2008a}.
For reference the centroid frequencies predicted by the Saclay
seismic model \citep{Turck-Chieze2001}, the M1 model
\citep{Zaatri2007}, and Model S \citep{Christensen1991} have been
included. The purple curve in Fig. \ref{figure[model echelle]} shows
the shape of the distribution of the model frequencies.

The frequencies predicted by the models for each individual mode are
spread over a reasonably large range, indicating the uncertainties
associated with solar modelling. However, there are definite gaps in
frequency where none of the models predict a mode should be
detected. Solar models would have to be changed beyond reasonable
expectations to account for any mode frequencies that lie outside
the predicted ranges. In the frequency ranges where no modes are
expected, we can set the prior probability that H0 is true to unity
thereby making it impossible for a prominent feature, which is
likely to be noise, to pass any statistical tests. We have,
therefore, defined an inverted top-hat prior that is equal to 0.5 in
the frequency ranges where, according to the models, the modes
should be observed and $p_0=1$ at all other frequencies. Such a
prior is shown in the top panel of Fig. \ref{figure[top hat]}. This
approach should severely limit the number of false detections that
are made.

\begin{figure}
\centering
  \includegraphics[width=0.4\textwidth, clip]{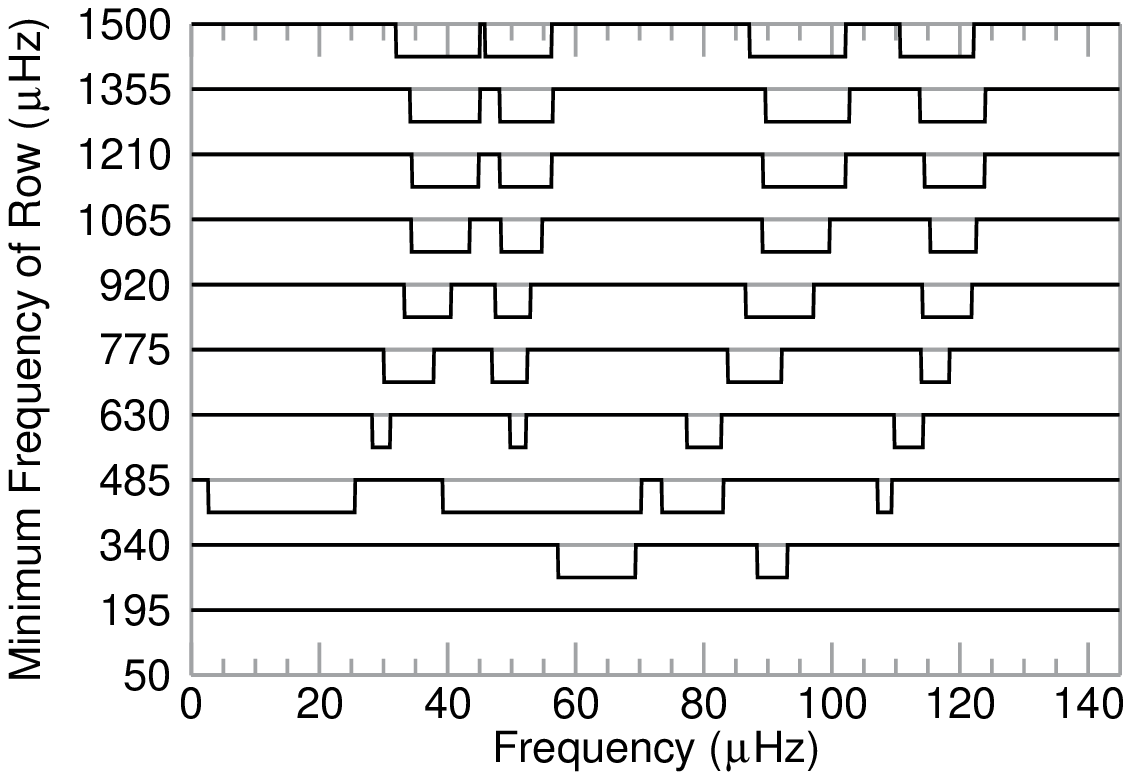}\\
  \includegraphics[width=0.4\textwidth, clip]{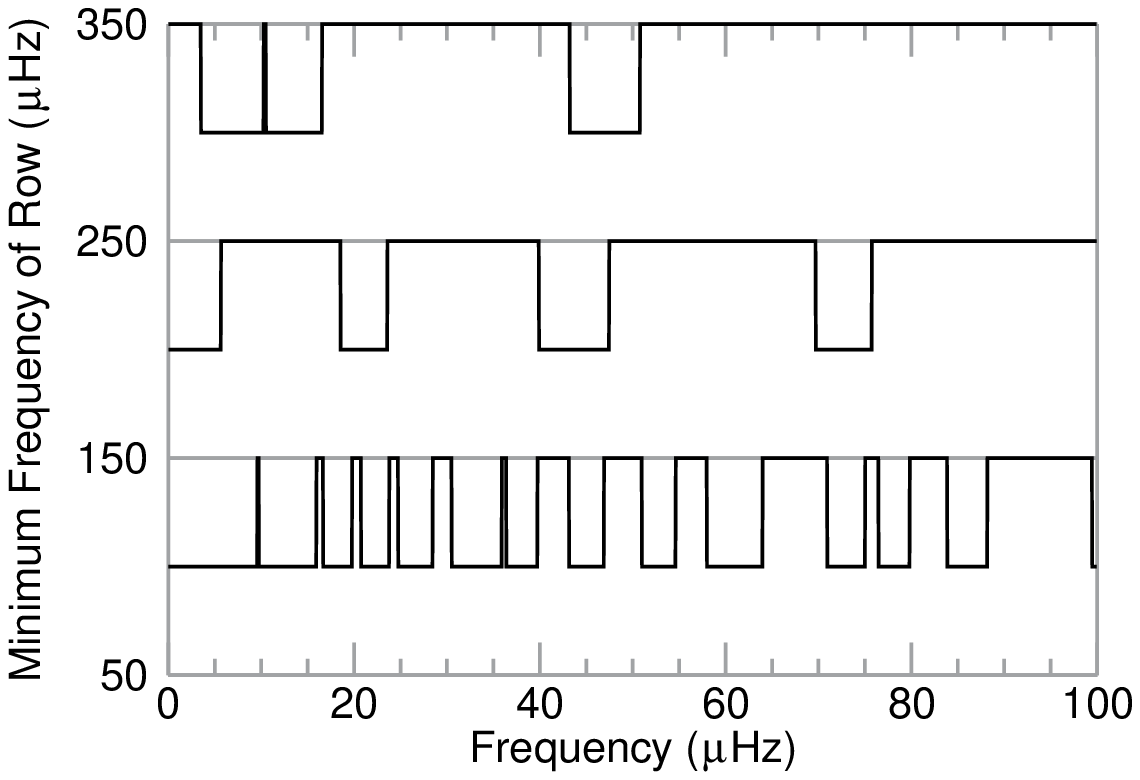}\\
  \caption{Top panel: An echelle diagram showing the variation of
  the prior probability with frequency for p~modes.
  Bottom panel: An echelle diagram showing the variation of the of
  the prior probability for g~modes. In each panel the prior probability was
  either 1.0, at frequencies where modes are not expected to be
  observed, or 0.5 at frequencies where modes are expected.}\label{figure[top
hat]}
\end{figure}

\subsection{A non-uniform prior that can be used to search for g~modes}
The sensitivity of g-mode frequency predictions to solar models is
of the order of 1\% \citep{Appourchaux2010}. This means, for
example, that at $100\,\rm\mu Hz$ the expected error on the
predicted mode frequencies is $1\,\rm\mu Hz$. We have used this
error estimate to produce two inverted top-hat prior distributions
for g~modes. For the first prior the centres of each hat are
determined by the mode frequencies predicted by the M1 model
\citep{Zaatri2007} and the width of each hat was 2 per cent of the
mode frequency i.e. $\pm1$ per cent. For the second prior, the width
of each hat was still 2 per cent of the mode frequency, however, we
have used the frequencies predicted by the Saclay solar model
\citep{Mathur2007}. Frequencies where modes are expected to be
observed have been given a prior probability that the H0 hypothesis
is true of 0.5 and elsewhere the prior is unity. The bottom panel of
Fig. \ref{figure[top hat]} shows the M1 model prior distribution,
where all visible components in Sun-as-a-star data (when $l+m$ is
even) have been plotted for $l=1$ and $l=2$ modes. We have not
plotted the Saclay model prior here as it is very similar to the M1
model prior. We have assumed that the different $m$ components are
still separated by $0.4\,\rm\mu Hz$.  As previously mentioned, the
results of \citet{Garcia2007} imply that this assumption may not be
valid for g~modes. In the absence of a definitive estimate for
g-mode splittings, we have chosen to maintain consistency with the
splitting value used when searching for p~modes. However, as new
information on the rotational splitting of g~modes becomes available
an updated uneven prior probability should be constructed. Notice
that despite the large number of g~modes that are present at low
frequencies the prior is not uniform with frequency and so, again,
this approach should limit the number of type~I false detections. It
is possible that an improved g-mode uneven prior probability could
be constructed by producing~models for g~modes that are similar to
those used to construct the p-mode uneven prior probability.

\section{Searching more than one data set using Bayesian
statistics}\label{section[Bayes stats]} 
\begin{figure}
\centering
  \includegraphics[width=0.4\textwidth, clip]{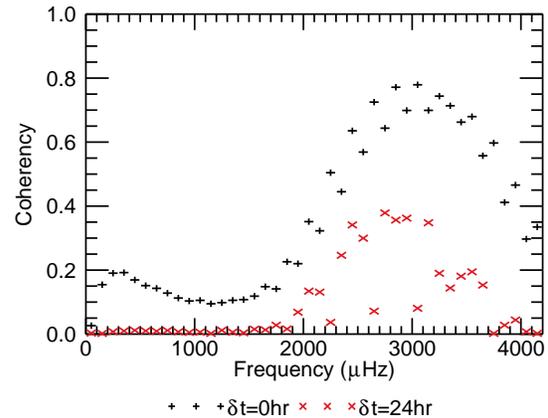}\\
  \caption{The coherency observed between BiSON and GOLF
  data when the GOLF data starts $24\,\rm h$ after the BiSON
data.}\label{figure[coherency]}
\end{figure}

As mentioned in Section
\ref{section[frequentist method]}, \citet{Broomhall2007} developed
the statistics necessary to allow two sets of contemporaneous data
to be compared using a frequentist approach. This has been done both
to lower the amplitude/power threshold levels and to improve the
reliability of any detection because if a prominent feature is found
in data observed by two different instruments it is more likely to
be solar in origin (although it should be noted that there may be
some consistent spacecraft signatures in the GOLF and MDI data).
However, it is worth mentioning that by comparing the data we are
limited by the signal-to-noise ratios of the worst instrument. For
example, there is a known instability problem with the MDI
instrument that causes low-frequency noise. Consequently, some mode
candidates that are only prominent in one data set have not been
highlighted here. This could increase the number of type~II errors
but, conversely, will limit the number of type~I errors, which we
consider to be more serious than type~II errors. In this paper, we
aim to use Bayesian statistics to compare multiple data sets. Since
contemporaneous BiSON, GOLF and MDI data will contain some common
noise (see Fig. \ref{figure[coherency]}), the probability that a
prominent spike, which is found in the same frequency bin in each
frequency-power spectrum, is due to noise is non-trivial to derive,
whether using a frequentist or Bayesian approach. We offer a
solution that avoids the need to consider the effect on the
statistics of the noise shared by each data set.

Fig. \ref{figure[coherency]} shows the coherency
\citep{Elsworth1994} of the BiSON and GOLF data when the GOLF time
series begins $24\,\rm h$ after the BiSON data. The separation in
start times means that the coherency of the noise observed between
the data is reduced to zero below $1500\,\rm\mu Hz$ and the same is
true for the other combinations of the BiSON, GOLF and MDI data.
Therefore, separating the start times of the data sets removes the
complication caused by the common noise. When no common noise is
shared by the two data sets under consideration the posterior
probability, $p(\textrm{H}00|x)$, that, if a feature appears in both
frequency-power spectra, H00 is correct, is given by
%%%%%%%%%%%%%%%%%%%%%%%%%%%%%%%%%%%%%%%%%%%%%%%%%%%%%%%%%%%%%%
\begin{equation}\label{equation[posterior 2 spectra]}
p(\textrm{H}00|x)=\left[1+\frac{(1-p_{0})}{p_{0}}\frac{p
(x_1|\textrm{H}1)p(x_2|\textrm{H}1)}{Mp(x_1|\textrm{H}0)
p(x_2|\textrm{H}0)}\right]^{-1},
\end{equation}
%%%%%%%%%%%%%%%%%%%%%%%%%%%%%%%%%%%%%%%%%%%%%%%%%%%%%%%%%%%%%%%%%
where $x_1$ is the power spectral density of a spike in the first
frequency-power spectrum and $x_2$ is the power spectral density of
a spike at exactly the same frequency in the second frequency-power
spectrum. When comparing two sets of non-contemporaneous data for
coincident prominent features we are assuming that the mode
signatures will still be coherent if we shift the start times and we
now show that this assumption is valid at low frequencies.

\subsection{The coherency of modes in
shifted spectra}\label{subsection[lifetime coherency]}

Simulations were performed to investigate how a modes's coherency
changes in non-contemporaneous data. The modes were simulated by
creating a damped oscillator that was re-excited randomly in time
\citep{Chaplin1997}. The power and damping time of each mode were
determined by fitting the BiSON data using a standard likelihood
maximization method \citep[e.g.][]{Chaplin1999}. Simulations were
performed for $l=1$ modes with $10\le n \le 22$ and a separate
simulation was performed for each mode. These modes are at higher
frequencies than the ones we are searching for in this paper.
However, they have a large S/N ratio and so their properties, such
as lifetime and power, can be determined more accurately. The
simulations were performed using two identical time series that
contained the signal from one simulated mode only.

%%%%%%%%%%%%%%%%%%%%%%%%%%%%%%%%%%%%%%%%%%%%%%%%%%%%%%%%%%%%%%%%%%%%%%%%%%%

\begin{figure}
\centering
  \subfigure{\includegraphics[width=0.4\textwidth,
clip]{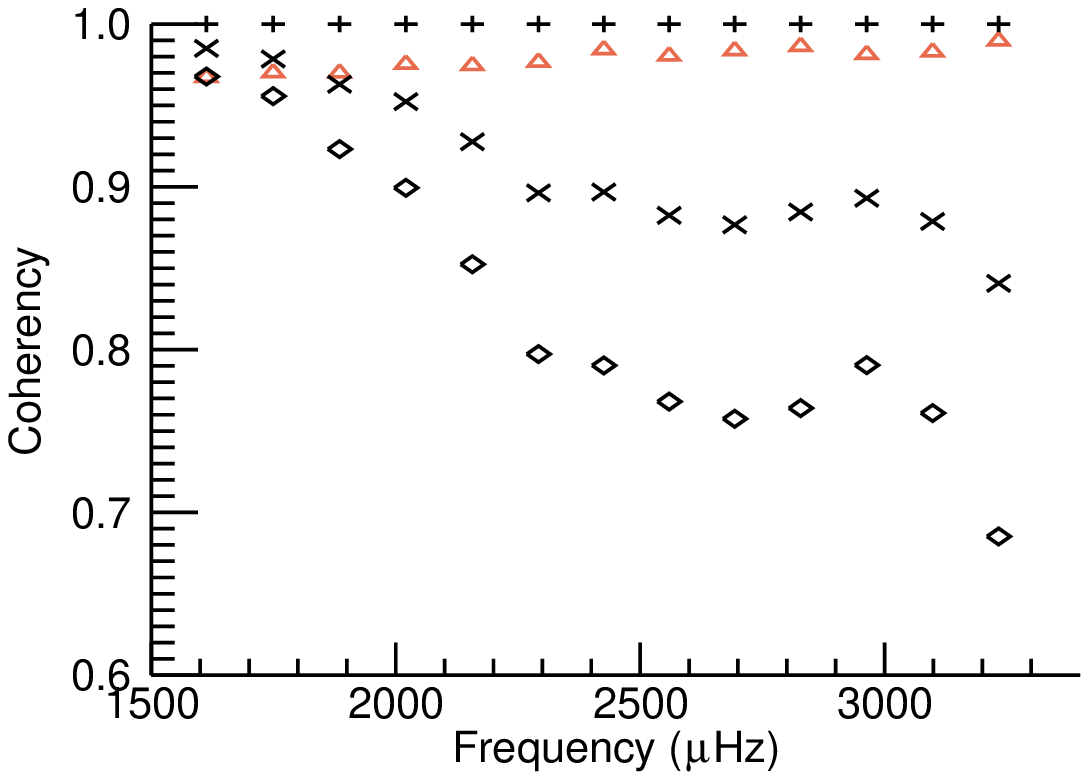}}\\
\vspace{0.5cm}
  \subfigure{\includegraphics[width=0.4\textwidth, clip]{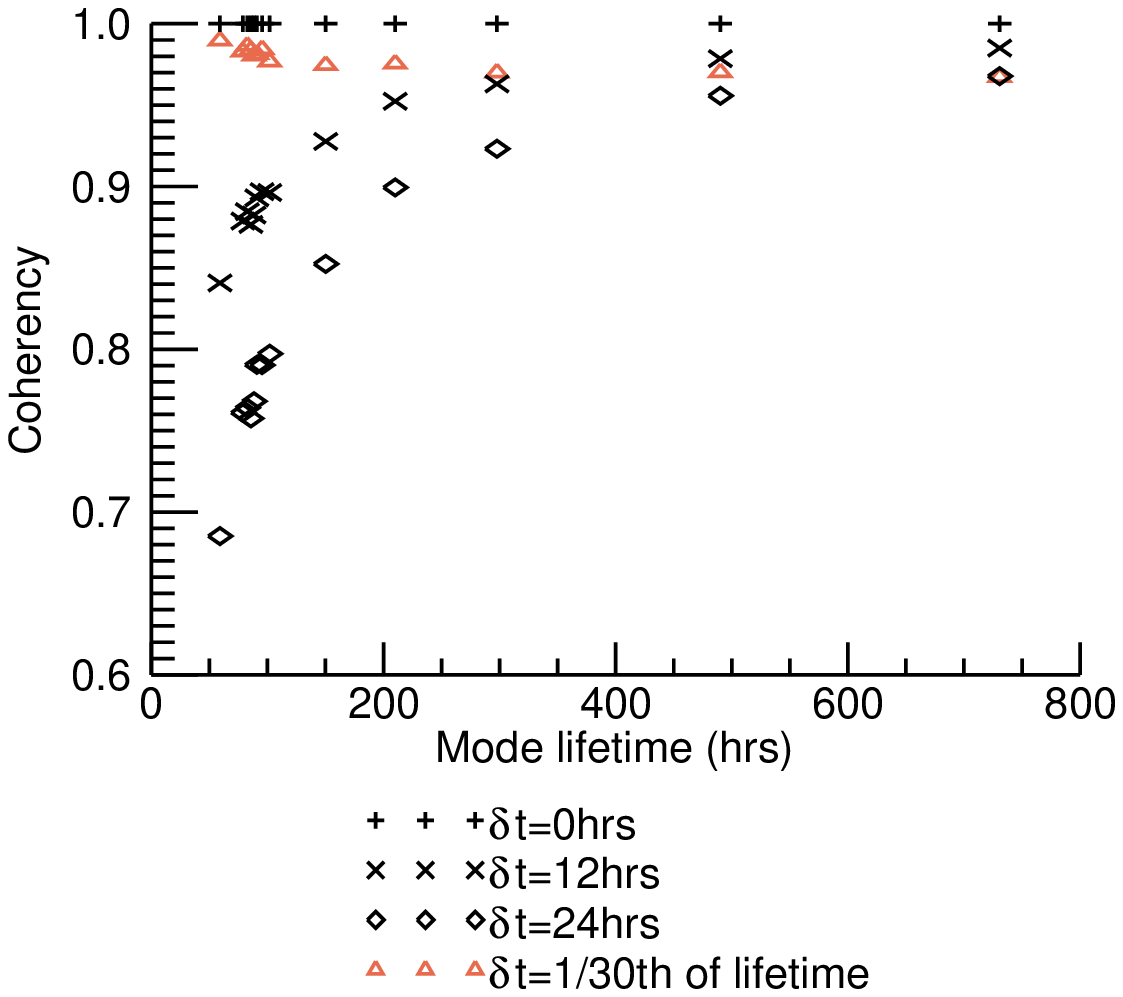}}\\
  \caption
  {Top panel: The effect of separating the start time of time series that
contain only the signal from a mode.
  The pairs of time series were created to contain one mode that has
  the same power and lifetime as is observed in $\sim8.5$yrs of
  BiSON data. Bottom panel: The coherency plotted as a function of lifetime.}
  \label{figure[simulated shift]}
\end{figure}

%%%%%%%%%%%%%%%%%%%%%%%%%%%%%%%%%%%%%%%%%%%%%%%%%%%%%%%%%%%%%%%%%%%%%%%%%%%%%

%%%%%%%%%%%%%%%%%%%%%%%%%%%%%%%%%%%%%%%%%%%%%%%%%%%%%%%%%%%%%%%%%%%%%%%%

\begin{figure}
\centering
  \includegraphics[width=0.4\textwidth,
clip]{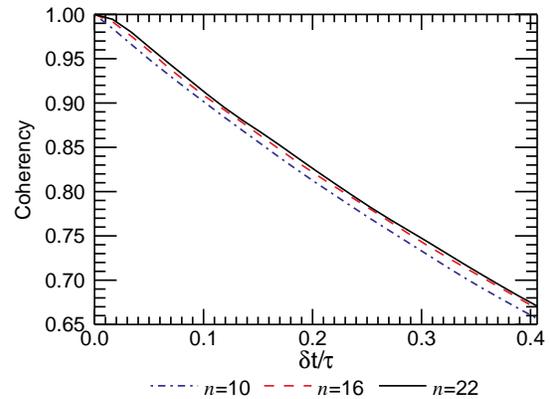}\\
  \caption{The variation in the coherency of three modes with $\delta t/\tau$.
  The results have been plotted for the $l=1$, $n=10$ mode, whose frequency
  is approximately $1612\,\rm\mu Hz$, the $l=1$, $n=16$ mode, whose frequency
  is approximately $2425\,\rm\mu Hz$ and the $l=1$, $n=22$ mode, whose frequency
  is approximately $3233\,\rm\mu Hz$.}\label{figure[sim shift coherency]}
\end{figure}

%%%%%%%%%%%%%%%%%%%%%%%%%%%%%%%%%%%%%%%%%%%%%%%%%%%%%%%%%%%%%%%%%%%%%%%%%%%%%%%%

The top panel of Fig. \ref{figure[simulated shift]} shows the
coherency \citep{Elsworth1994} of the modes when the start times of
the simulated time series were shifted so that they were separated
by $\delta t=0\,\rm h$ (black plus signs), $\delta t=12\,\rm h$
(black crosses) and $\delta t=24\,\rm h$ (black diamonds). The
coherency of two unshifted identical sets of data is always unity,
independent of the power and width of the simulated mode. The top
panel of Fig. \ref{figure[simulated shift]} shows that the effect on
the coherency of a mode of separating the start times of two data
sets is frequency dependent. More specifically (bottom panel of Fig.
\ref{figure[simulated shift]}) the magnitude of the decrease in
coherency is lifetime dependent. The coherency of a low-frequency
mode, with a long lifetime, is reduced by less than the coherency of
a higher-frequency mode, with a short lifetime. This is because, as
the time shift represents a smaller proportion of a low-frequency
modes' lifetime, the mode experiences less damping over the duration
of the shift. The magnitude of the decrease in coherency increases
as the length of the shift in start times increases relative to the
mode lifetime. However, is the decrease in coherency observed by
modes below $1500\,\rm\mu Hz$ significant when a time shift of
$24\,\rm h$ is introduced?

Modes with a frequency of about $1500\mu$Hz have a lifetime of
approximately 1 month and so $24\,\rm h$ is equivalent to $1/30$th
of the mode's lifetime. Therefore, for each of the modes plotted in
Fig. \ref{figure[simulated shift]}, the start times of the two
simulated time series were separated by one-thirtieth of the mode's
lifetime. The triangle symbols in the top panel of Fig.
\ref{figure[simulated shift]} shows that this only decreases the
coherency by a very small amount at all frequencies. Fig.
\ref{figure[sim shift coherency]} plots the variation in the
coherency with $\delta t/\tau$ for three $l=1$ modes. The values of
coherency plotted in Fig. \ref{figure[sim shift coherency]} were
determined using the same simulated spectra as were used to produce
Fig. \ref{figure[simulated shift]}. Fig. \ref{figure[sim shift
coherency]} shows that if $\delta t/\tau$ is small, the coherency
decreases by only a small amount. Notice that size of the decrease
in coherency shows some dependence on $n$. However, this dependence
is small and is likely to occur because the power of the modes
decreases with $n$. We conclude that, if the start times of two time
series are separated by $24\,\rm h$ the low-frequency modes will
still be coherent and so the data can be compared to search for
low-frequency oscillations.

Note that it is possible to separate the start times of the time
series by less than $24\,\rm h$ and still significantly reduce the
amount of common noise shared by the data. However, we are dealing
with long data sets and so $24\,\rm h$ is only a small fraction of
the total observation time. Therefore a shift of $24\,\rm h$ was
used to ensure that there was as little common noise present as
possible.

A table summarizing the main assumptions made with both the
frequentist and Bayesian approaches can be found in Appendix
\ref{section[appendix]}. We now describe the results of using both
the frequentist approach and Bayesian statistics to search BiSON,
GOLF and MDI data.

\section{Results of searching pairs of data sets for low-frequency p~modes and g~modes}\label{section[results]}

\subsection{Observational data} In this paper we made use of contemporaneous
BiSON, GOLF and MDI time series. Each data set consisted of
$3071\,\rm d$ of Sun-as-a-star Doppler velocity observations, made
between 1996 April 20 and 2004 September 15, and so the data span
most of solar activity cycle 23. We have used data sets that extend
only as far as 2004 to maintain consistency with
\citet{Broomhall2007}. The BiSON data were processed in the manner
described by \citet{Appourchaux2000} and \citet{Chaplin2002}. The
GOLF data were processed in the manner described by
\citet{Garcia2005}, and the MDI-Luminosity Oscillations Imager (LOI)
proxy data were processed in the manner described by
\citet{Scherrer1995}. When considered individually each time series
is stored on a different cadence: BiSON data are stored on a
$40\,\rm s$ cadence, GOLF data are stored on a $20\,\rm s$ cadence
and MDI data are stored on a $60\,\rm s$ cadence. Each time series
was, therefore, re-binned to a common cadence of $120\,\rm s$. Hence
the time series contained 2211 120 samples, giving a bin width in
the frequency domain of $0.004\,\rm \mu Hz$. It is possible that
some aliasing could be present in the data due to the cut-off
frequency of $4167\,\rm\mu Hz$. However, no aliasing was found in
the frequency range searched here i.e. between 50 and $1500\,\rm\mu
Hz$. The duty cycle of the BiSON time series was 78.6 per cent; the
duty cycle of the GOLF time series was 93.4 per cent and the duty
cycle of the MDI time series was 91.3 per cent.

\subsection{Using a frequentist approach}
The frequentist statistics described in Section
\ref{section[frequentist method]} and \citet{Broomhall2007} were
used to search contemporaneous BiSON, GOLF and MDI
frequency-amplitude spectra for low-frequency p~modes, g~modes and
mixed modes. Amplitude threshold levels were calculated for a 1 per
cent probability of detecting a prominent feature by chance in
$100\,\rm\mu Hz$. The spectra were searched for prominent spikes or
patterns of spikes found in the same frequency bin or bins in any
two of the frequency-amplitude spectra.

If the mode signal is not commensurate with the window function of
the observations the maximum observed amplitude of the mode can be
diminished. The `bin-shifting' strategy of \citet{Chaplin2002} was
therefore applied in an attempt to circumvent this problem.

%%%%%%%%%%%%%%%%%%%%%%%%%%%%%%%%%%%%%%%%%%%%%%%%%%%%%%%%%%%%%%%%%%%%%%%%%%%%%%%

\begin{figure}
\centering
  \includegraphics[width=\columnwidth, clip]{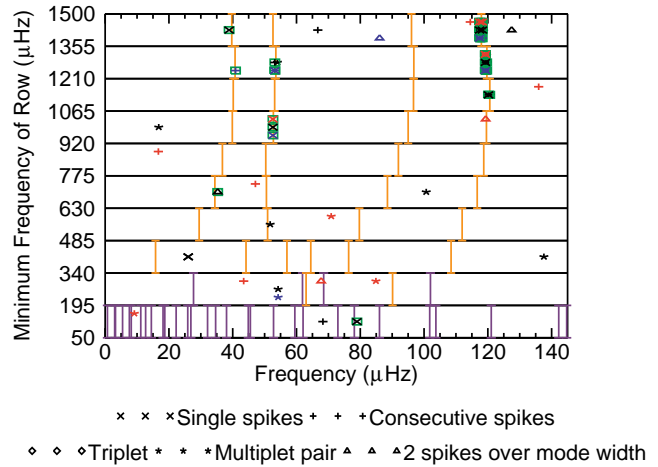}\\
 % \vspace{0.3cm}
 % \includegraphics[width=0.78\textwidth, clip]{legend_echelle1.eps}\\
  \caption{An echelle plot, modulo $145\,\rm \mu Hz$, marking locations in
frequency of
  occurrences uncovered by the test searches. Locations in
  frequency where spikes, or patterns of spikes, were found
  in the same bin, or bins, in BiSON and GOLF frequency-amplitude spectra at
levels
  sufficient to record $P \le 1$\,per cent are marked by the black
  symbols in the middle of each row. A different symbol has been used
  for each test (see figure legend). We have also recorded prominent
  spikes or patterns of spikes found by comparing either the BiSON
  and MDI frequency-amplitude spectra
  (red symbols at top of each row) or the GOLF and MDI frequency-amplitude
spectra (blue symbols at bottom of
  each row). Symbols surrounded by a green square represent
  the prominent occurrences listed in Table \ref{table[coincidence search
  results]}.
  The orange vertical lines mark locations of the frequencies of p~modes
with $l=0-3$ predicted by the Saclay
  seismic model \citep{Turck-Chieze2001}. The vertical purple lines
  mark locations of the central frequencies of g~modes with $l=1-2$
predicted by the M1
  model \citep{Zaatri2007}.}
  \label{figure[results echelle]}
  \end{figure}

%%%%%%%%%%%%%%%%%%%%%%%%%%%%%%%%%%%%%%%%%%%%%%%%%%%%%%%%%%%%%%%%%%%%%%%%%%%%%%%

Fig. \ref{figure[results echelle]} shows visually, in the form of an
echelle diagram, the locations in frequency where prominent spikes
or patterns of spikes were observed. The entire range that was
searched ($50-1500\,\rm\mu Hz$) has been split into strips of
$145\,\rm\mu Hz$. These strips have then been placed one above
another. The bottom left-hand corner of the plot represents
$50\,\rm\mu Hz$ and the top right-hand corner of the plot represents
$1500\,\rm\mu Hz$. A repeat frequency of $145\,\rm\mu Hz$ was chosen
in preference to the $100\,\rm\mu Hz$ slice on which the searches
were performed because at low frequencies consecutive overtones of
the low-$l$ modes are separated in frequency by approximately this
amount. At higher frequencies the spacing between modes with the
same $l$ and adjacent $n$ is $\sim135\,\rm\mu Hz$, however, this
separation increases at low frequencies (low $n$). The choice of
repeat frequency means that the majority of the predicted p-mode
frequencies are arranged in four near-vertical strips, one for each
$l$ between 0 and 3.

Artifacts, positioned in frequency at overtones of the
$11.57\,\rm\mu Hz$ diurnal frequency, may be visible in
frequency-amplitude spectra because of signatures of either the
window function (for the BiSON data) or the spacecraft operation
(for the GOLF and MDI data). To prevent any of these artefacts being
confused with mode candidates, any detections found to be closer
than $0.3\,\rm\mu Hz$ to a daily harmonic frequency were discounted.
A distance of $0.3\,\rm\mu Hz$ is sufficient to remove any false
detections due to the daily harmonics as their frequencies are well
defined and their peaks have a width of less than $0.3\,\rm\mu Hz$.
The MDI data contain harmonics of $52.125\,\rm\mu Hz$ due to beats
between the spacecraft timing system and the instrument sampling
rate. Any detections found when the MDI data were searched that lay
within $0.3\,\rm\mu Hz$ of these harmonic frequencies were also
discounted.

%%%%%%%%%%%%%%%%%%%%%%%%%%%%%%%%%%%%%%%%%%%%%%%%%%%%%%%%%%%%%%%%%%%
\begin{table*}
\caption{Candidates found to be closer than $1\mu$Hz to the
predicted frequencies of p~modes from the Saclay seismic model and
the predicted frequencies of g~modes from the M1 model. $P$ is the
probability that, assuming the data contains only noise, a
coincident prominent feature is observed at least once in $100\,\rm
\mu Hz$. The last column of this table indicates whether the modes
were also detected using a Bayesian approach (see Sections
\ref{section[results bayes single]} and \ref{section[results bayes
smooth]}).}\label{table[coincidence search results]}
\begin{tabular}{cccccccc}
  \hline
  \noalign{\smallskip}
  \small{\emph{l}} &\small{\emph{n}} & \small{\emph{m}} & \small{Frequency} &
  \small{Probability} & \small{Number} & Distance & Detected\\
   &  &  & \small{($\mu$Hz)} &
  \small{($P$)} & \small{of tests} & from model & with the\\
   &  &  &  & & passed & frequency  & Bayesian\\
   &  &  &  &  &  & ($\mu$Hz) & approach\\
  \noalign{\smallskip}
  \hline
  \noalign{\smallskip}
  1 & -4 & $+1$ & $128.951\pm0.002$ & $8.7\times10^{-3}$ & 1 & 0.421$^a$ &
n\vspace{2mm}\\
  2 & 3 & $-2$ & $664.515\pm0.002$ & $6.7\times10^{-3}$ & 1 & & n\\
  2 & 3 & $0$ & $665.284\pm0.002$ & $6.7\times10^{-3}$ & 1 & 0.880 & n\\
  2 & 3 & $+2$ & $666.128\pm0.002$ & $6.7\times10^{-3}$ & 1 & & n\vspace{2mm}\\
  0 & 6 & $\quad0\quad$ & $972.613\pm0.002$ & $3.9\times10^{-5}$ & 1 & 0.132 &
y\vspace{2mm}\\
  1 & 7 & $-1$ & $1185.196\pm0.005$ & $1.9\times10^{-3}$ & 4 & & y\\
  1 & 7 & $+1$ & $1185.981\pm0.005$ & $3.0\times10^{-3}$ & 1 & & n\\
  & & mean & $1185.589\pm0.004$ & & & 0.027 & \vspace{2mm}\\
  2 & 7 & 0 & $1250.893\pm0.007$ & $6.3\times10^{-3}$ & 1 & 0.159 & n\vspace{2mm}\\
  0 & 8 & 0 & $1263.205\pm0.007$ & $2.1\times10^{-4}$ & 2 & 0.319 & y\vspace{2mm}\\
  1 & 8 & $-1$ & $1329.236\pm0.005$ & $2.7\times 10^{-11}$ & 4 & & y\\
  1 & 8 & $+1$ & $1330.037\pm0.007$ & $1.7\times10^{-3}$ & 1 & & y\\
  & & mean & $1329.637\pm0.004$ & & & 0.060 & \vspace{2mm}\\
  2 & 8 & $-2$ & $1393.871\pm0.007$ & $4.0\times10^{-4}$ & 1 & 0.036$^b$ &
y\vspace{2mm}\\
  1 & 9 & $-1$ & $1472.432\pm0.008$ & $3.8\times 10^{-11}$ & 4 & & y\\
  1 & 9 & $+1$ & $1473.269\pm0.009$ & $3.7\times 10^{-8}$ & 4 & & y\\
  & & mean & $1472.851\pm0.006$ & & & 0.122 & \\
  \noalign{\smallskip}
  \hline
\end{tabular}
\flushleft{\small$^a$Difference with model frequency assumes
$m=\pm1$ components lie $\pm0.4\,\rm \mu Hz$ from central
frequency.\\ $^b$Difference with model frequency assumes $m=-2$
component lies $-0.8\,\rm \mu Hz$ from the central frequency.}
\end{table*}
%%%%%%%%%%%%%%%%%%%%%%%%%%%%%%%%%%%%%%%%%%%%%%%%%%%%%%%%%%%%%%%%%%%%%%%%%

Detections were only considered as possible mode candidates if they
were positioned within $1\,\rm\mu Hz$ of a predicted mode frequency.
A maximum separation between a detection and a model frequency of
$1\,\rm\mu Hz$ was allowed to account for any uncertainties in the
model frequencies, any shifts due to the effects of magnetic fields,
and because the power of some modes is spread across more than one
frequency bin. It should be noted that Figs \ref{figure[model
echelle]} and \ref{figure[top hat]}, which show the frequencies
predicted by different solar models and the uneven priors used in
the Bayesian approach, suggest that the allowed difference with the
model should be larger than $\pm1\,\rm\mu Hz$. However, it was felt
that an increase in the allowed difference between the model and the
observed frequencies would significantly increase the number of
type~I false detections that were considered as mode candidates. Any
detections that lie within $1\,\rm\mu Hz$ of a model frequency are
surrounded by a green square in Fig. \ref{figure[results echelle]}
and are listed in Table \ref{table[coincidence search results]}. We
have compared the observed mode candidates with both the Saclay
seismic model \citep{Turck-Chieze2001, Mathur2007} and the M1 model
frequencies \citep{Zaatri2007}. However, we have not plotted all of
these frequencies in Fig. \ref{figure[results echelle]} as both
models predict similar frequencies. In cases where the outer
components of a rotationally split mode appear to have been
uncovered, the mean of the component frequencies has been
determined, in order to give an estimate of the centroid frequency
of the mode. Estimates of the uncertainties in frequency were
calculated in the manner described by \citet{Chaplin2002}.

In the p-mode frequency range all of the listed candidates
correspond to previously claimed detections \citep[see
e.g.][]{Toutain1998, Bertello2000, Garcia2001, Chaplin2002,
Broomhall2007, Salabert2009}. One of the detections in the g-mode
range lies close to the predicted frequencies. To improve confidence
in the detections we required that each mode candidate passed more
than one of the statistical tests, as it has been determined that
this requirement significantly reduces the number of false
detections made. As can be seen from Table \ref{table[coincidence
search results]} this discounts the g-mode candidate and some of the
detections in the p-mode range. For comparison purposes the last
column of Table \ref{table[coincidence search results]} indicates
whether the candidates were also detected when Bayesian statistics
were used (see Sections \ref{section[results bayes single]} and
\ref{section[results bayes smooth]}).

Several multiplet detections lie close to the predicted mode
frequencies but are not highlighted by a green square in Fig.
\ref{figure[results echelle]}. This is because not all of the
detected components could be associated with the predicted
frequencies of the components of one mode. For example, a triplet
was detected in the BiSON and MDI data at $\sim1039\,\rm\mu Hz$.
However, the mode closest to the detection is the $l=1$, $n=6$ mode,
which only has two visible components $(m=\pm1)$. Since only two of
the three detected spikes could be associated with model mode
frequencies this was not counted as a mode candidate. Similarly the
multiplet candidate at $\sim537\,\rm\mu Hz$ is closest to the $l=0$,
$n=3$ mode, which cannot be observed as a multiplet (as $l=0$).

In Fig. \ref{figure[results echelle]} we can see that coincident
multiplet detections are made at $\sim249\,\rm\mu Hz$ in the BiSON
and GOLF data and the GOLF and MDI data, which do not correspond to
a predicted mode frequency. One of the detected components is at the
same frequency in each detection (at $248.474\,\rm\mu Hz$). This
component is particularly prominent in the GOLF data. The second
component observed in the BiSON and GOLF data is not at the same
frequency as the second component that is observed in the GOLF and
MDI data. Therefore, these detections are not truly coincident.

It is apparent from Fig. \ref{figure[results echelle]} that the
analysis has uncovered several occurrences of $P\le1$ per cent that
lie well away from the predicted mode frequencies. The number of
these detections exceeds the number expected $(<2)$ given the range
in frequencies searched, the threshold probability, and the number
of statistical tests performed. This is most likely to be due to the
misleading significance levels assigned when adopting a frequentist
approach. It could also indicate that the statistical distribution
in the noise is slightly different to the Gaussian distribution,
which was assumed when deriving the statistical tests. It is also
worth noting that given the large number of g~modes that can
theoretically be detected, particularly at very low frequencies, it
is hardly surprising that some of the detections lie close to
predicted mode frequencies.

\subsubsection{Searching for g~mode peaks in contemporaneous data} As
previously mentioned the results obtained by \citet{Garcia2007}
could imply that g~modes may have lifetimes of the order of months,
not years. If this is true they will have a resolved width in a
frequency-amplitude spectrum and so we have searched the
contemporaneous, very low-frequency $(\nu<350\,\rm\mu Hz)$ BiSON,
GOLF and MDI data for evidence of prominent peaks with various
widths. We have searched for clusters containing two, three, four
and five spikes. We allow a cluster to be spread over twice the
width of the mode and took the widths of the modes to be the number
of bins covered if the mode lifetimes were 1, 2, 3 and 4 months,
respectively. However, no statistically significant peaks were found
within $4\,\rm\mu Hz$ of the model g-mode frequencies.

%%%%%%%%%%%%%%%%%%%%%%%%%%%%%%%%%%%%%%%%%%%%%%%%%%%%%%%%%%%%%%%%%%%%%%

\begin{figure}
\centering
  \includegraphics[angle=90, width=0.4\textwidth]{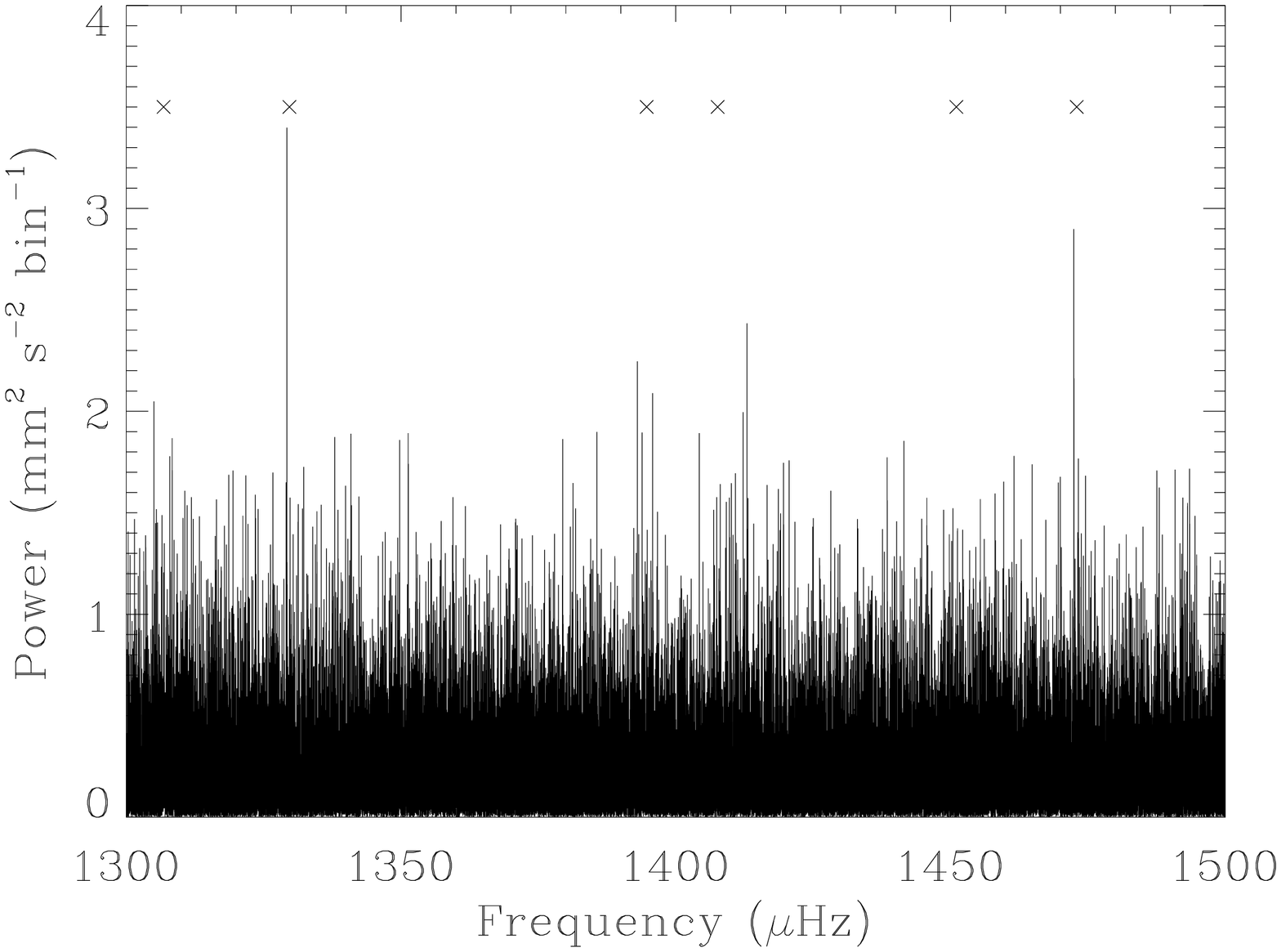}\\
  \vspace{2mm}
  \includegraphics[angle=90, width=0.4\textwidth]{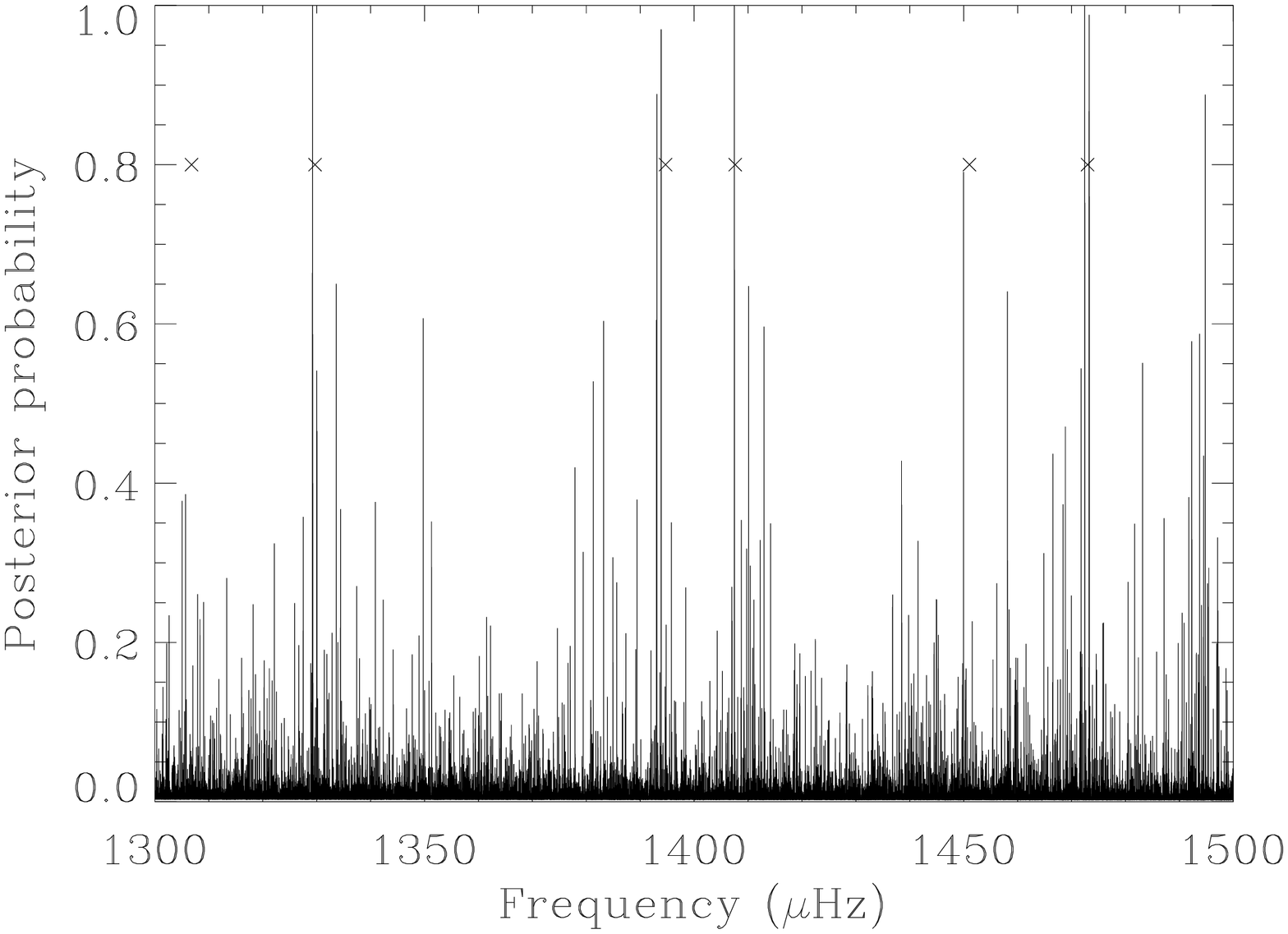}\\
   \vspace{2mm}
  \includegraphics[angle=90, width=0.4\textwidth]{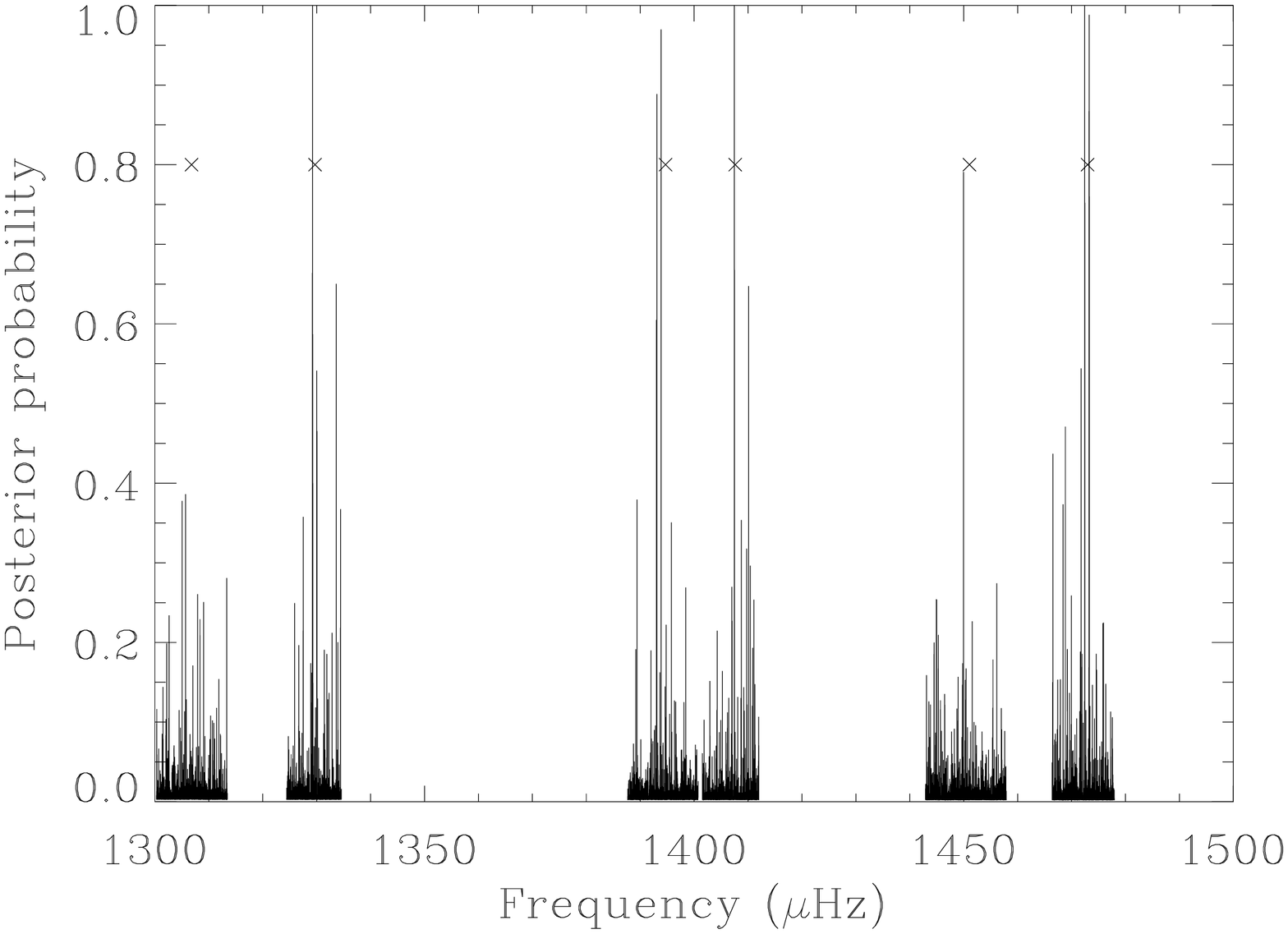}\\
  \caption{Top panel: a section of the BiSON frequency-power
  spectrum. Middle panel: the posterior probability observed when
  BiSON and GOLF data are searched for p~modes. The BiSON data was
  shifted to start $24\,\rm h$ after the GOLF data. A uniform
  prior probability of $p_0=0.5$ was used. Bottom panel: The
  posterior probability observed when BiSON and GOLF data were
  searched for p~modes. However, in this panel a the inverted
  top-hat prior described in Section \ref{section[prior]} was used.
  }\label{figure[posterior prob]}
\end{figure}

%%%%%%%%%%%%%%%%%%%%%%%%%%%%%%%%%%%%%%%%%%%%%%%%%%%%%%%%%%%%%%%%%%%%%%%%%%

\subsection{Results of searching the data using Bayesian
techniques}\label{section[results bayes single]} The statistics
outlined in Sections \ref{section[Bayes intro]}-\ref{section[Bayes
stats]} were used to search BiSON, GOLF and MDI data in pairs with a
Bayesian approach, respectively. The start times of each pair of
time series were separated by $24\,\rm h$ so that only a minimal
amount of common noise was present in the data. The `bin-shifting'
strategy of \citet{Chaplin2002} was again applied. Equation
(\ref{equation[posterior 2 spectra]}) was used to determine the
posterior probability, $p(\textrm{H}00|x)$, where initially we took
the height of the p~modes, $H$, to be given by the extrapolation
described in \citet{Broomhall2008} (see Section
\ref{section[amplitude range]}). We have assumed that g~mode
lifetimes are sufficiently long for all of their power to be
contained in one bin and the maximum g-mode height was taken to be
$9\,\rm mm^2\,s^{-2} bin^{-1}$ at all frequencies. The frequency
range of $250-1500\,\rm\mu Hz$ was searched for solar p~modes and
the frequency range of $50-350\,\rm\mu Hz$ was searched for g~modes.

The top panel of Fig. \ref{figure[posterior prob]} shows the BiSON
spectrum over the range $1300\,\rm\mu Hz$ to $1500\,\rm\mu Hz$. The
crosses above the spectrum show the p-mode frequencies predicted by
the Saclay seismic model \citep{Turck-Chieze2001}. As can be seen,
it is difficult to distinguish possible mode candidates from the
background noise. The middle panel of Fig. \ref{figure[posterior
prob]} shows $[1-p(\textrm{H}00|x)]$, for observing a prominent
feature in the BiSON and GOLF data, calculated with a uniform prior
of $p_0(\nu)=0.5$. As $[1-p(\textrm{H}00|x)]$ approaches unity the
more likely we are to reject the H00 hypothesis that the
frequency-power spectrum contains noise only. Four mode candidates
are clearly visible in this plot, at $\sim1330\,\rm\mu Hz$, $\sim
1393\,\rm\mu Hz$ and two at $\sim 1472\,\rm\mu Hz$. All four
candidates lie close to predicted mode frequencies. The bottom panel
shows the posterior probability over the same range of frequencies,
only this time the top-hat prior probability described in Section
\ref{section[prior]} was used. All of the mode candidates are
allowed by the uneven prior.

\begin{figure}
\centering
  \includegraphics[width=\columnwidth, clip]{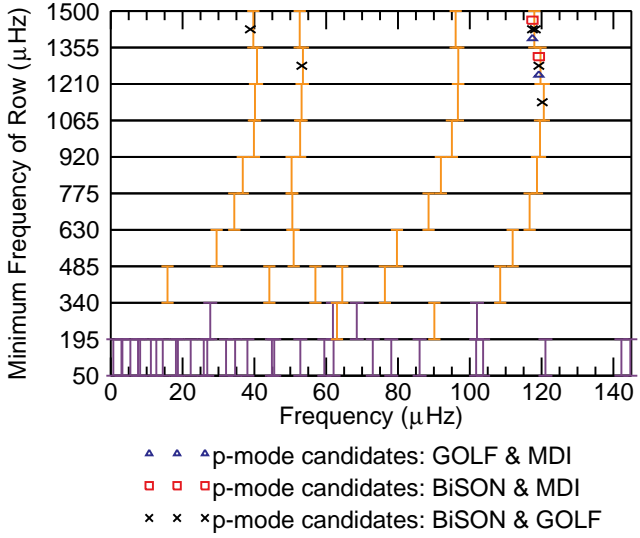}\\
  \caption{An echelle plot, modulo $145\,\rm \mu Hz$, marking locations in
frequency of
  occurrences uncovered by the Bayesian test searches. Realistic estimates
  of the upper limit on the modes powers were used to calculate the
  Bayesian probability, $p(\textrm{H}0|x)$.  The different
  symbols represent the different pairings of data that were searched (see
legend).}\label{figure[Bayes echelle]}
\end{figure}

Fig. \ref{figure[Bayes echelle]} is an echelle diagram showing the
results of searching the BiSON, GOLF and MDI data using a Bayesian
approach. Mode candidates have been plotted if
$p(\textrm{H}00|x)\le0.01$. Therefore, at the plotted frequencies,
we reject the H00 hypothesis that the data contains noise only, in
favour of the H11 hypothesis which states that there is some signal
in the data. The non-uniform prior described in Section
\ref{section[prior]} was applied and resulted in one p-mode
candidate being discounted. This candidate was detected in the
BiSON-MDI combination and the GOLF-MDI combination. The frequency of
the candidate corresponds to a spacecraft frequency that is very
prominent in the MDI frequency-power spectrum. Therefore, the fact
that this candidate is discounted is a good example of how
advantageous the uneven prior can be. Very few detections are made
in the p-mode frequency range. However, all of the candidates that
are detected correspond to mode frequencies. This implies that the
Bayesian approach can successfully discriminate between signal and
noise. No candidates are found in the g-mode range.

\begin{figure}
\centering
  \includegraphics[width=\columnwidth, clip]{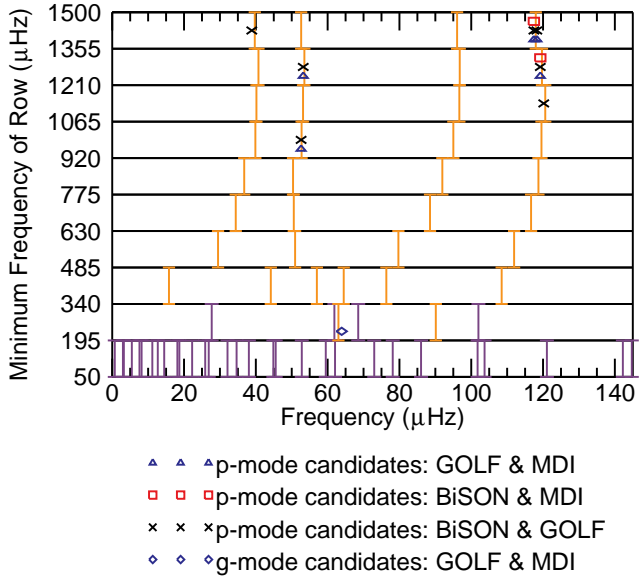}\\
  \caption{An echelle plot, modulo $145\,\rm \mu Hz$, marking locations in
frequency of
  occurrences uncovered by the Bayesian test searches. The maximum
  power, $H$, was overestimated when calculating $p(\textrm{H}0|x)$. The
different
  symbols represent the different pairings of data that were searched (see
legend).}
  \label{figure[Bayes echelle 1500]}
\end{figure}

Fig. \ref{figure[Bayes echelle 1500]} is an echelle diagram that
again shows the results of using Bayesian methods to search for
solar oscillations. The only difference between Figs
\ref{figure[Bayes echelle]} and \ref{figure[Bayes echelle 1500]} is
that larger maximum heights, $H$, have been used when determining
the posterior probability. When searching for the p~modes we set $H$
to be the height a mode at $1500\,\rm\mu Hz$ is expected to have.
When searching for g~modes the maximum height was taken to be
$36\,\rm mm^2\,s^{-2} bin^{-1}$, which is four times larger than the
heights assumed when producing Fig. \ref{figure[Bayes echelle]} (see
Section \ref{section[amplitude range]} for details). More mode
candidates are detected in the p-mode range than were found when the
lower heights were used (see Fig. \ref{figure[Bayes echelle]}).
Furthermore, the $p(\textrm{H}00|x)$ is lower when the larger values
of $H$ were used for all of the candidates that are plotted in both
Figs \ref{figure[Bayes echelle]} and \ref{figure[Bayes echelle
1500]}. This could indicate that we were correct to overestimate the
upper limit of $H$ when determining $p(x|\rm H11)$.

Fig. \ref{figure[Bayes echelle 1500]} shows that one g-mode
candidate was detected. Even though this candidate is close to a
predicted p-mode frequency ($l=0$, $n=1$) it has been classified as
a g-mode candidate as it was detected when the value for $H$ was
assumed to be $36\,\rm mm^2\,s^{-2} bin^{-1}$, i.e. $H$ was set
using the g-mode models. However, we do not rule out the possibility
that this detection is a p~mode. A peak in this frequency region is
mentioned by \citet{Turck-Chieze2004}, who looked at GOLF data only.

Table \ref{table[Bayesian results]} lists the mode candidates that
are shown in Fig. \ref{figure[Bayes echelle 1500]} along with the
minimum observed posterior probability, $p(\textrm{H}00|x)$. The
value of $p(\textrm{H}00|x)$ quoted in Table \ref{table[Bayesian
results]} is dependent on which of the two sets of data were
considered. The $l=1$, $n=9$, $m=-1$ mode candidate at
$1472.433\,\rm\mu Hz$ was detected at a significant level in all
three combinations of data (see Fig. \ref{figure[Bayes echelle
1500]}). It was most prominent in the BiSON-GOLF combination, when
$p(\textrm{H}00|x)=1.6\times10^{-7}$. When this candidate was
observed in the GOLF-MDI combination
$p(\textrm{H}00|x)=2.3\times10^{-5}$ and when it was observed in the
BiSON-MDI combination $p(\textrm{H}00|x)=1.4\times10^{-3}$. When a
candidate was detected in more than one combination of data sets the
minimum value of $p(\textrm{H}00|x)$ is included in Table
\ref{table[Bayesian results]}. In fact all of the included values of
$p(\textrm{H}00|x)$ were taken from the BiSON-GOLF combination,
except for the g~mode candidate, which was observed in the GOLF-MDI
combination. However, unlike the $l=1$, $n=9$, $m=-1$ candidate,
some mode candidates were observed to be more prominent in the
BiSON-MDI combination than in the GOLF-MDI combination. For example,
the $l=1$, $n=8$, $m=-1$ mode candidate was found to have a
posterior probability of $7.4\times10^{-8}$ when observed in the
BiSON-MDI combination and $6.6\times10^{-7}$ when observed in the
GOLF-MDI combination. It should be noted that all of the results
described in this section were dependent on the assumptions made
when determining $p(H00|x)$. Different initial assumptions may have
led to different results.

\begin{table}
\caption{Candidates found using Bayesian
statistics.}\label{table[Bayesian results]}
\begin{tabular}{cccccc}
  \hline
  \noalign{\smallskip}
  \small{\emph{l}} &\small{\emph{n}} & \small{\emph{m}} & \small{Frequency} &
  \small{Posterior} &  Distance \\
   &  &  & \small{($\mu$Hz)} &
  \small{probability} & from model \\
   &  &  &  & \small{$[p(\textrm{H}00|x)]$} & frequency  \\
   &  &  &  &  & ($\mu$Hz)\\
   \noalign{\smallskip}
   \hline
   \noalign{\smallskip}
  2 & -2 & $\quad+2\quad$ & $258.916\pm0.002$ & $7.0\times10^{-3}$ & 1.236 \vspace{2mm}\\
  0 & 6 & $\quad0\quad$ & $972.613\pm0.002$ & $4.2\times10^{-3}$ & 0.132 \vspace{2mm}\\
  1 & 7 & $-1$ & $1185.197\pm0.005$ & $5.1\times10^{-4}$ & 0.256$^a$\vspace{2mm}\\
  0 & 8 & 0 & $1263.209\pm0.007$ & $6.2\times10^{-4}$ &  0.397 \vspace{2mm}\\
  1 & 8 & $-1$ & $1329.235\pm0.005$ & $4.5\times 10^{-13}$ & 0.273$^a$ \vspace{2mm}\\
  2 & 8 & $-2$ & $1393.872\pm0.007$ & $3.6\times10^{-3}$ & 0.336$^b$ \vspace{2mm}\\
  1 & 9 & $-1$ & $1472.433\pm0.008$ & $1.6\times 10^{-7}$ & \vspace{2mm}\\
  1 & 9 & $+1$ & $1473.270\pm0.009$ & $1.9\times 10^{-4}$ & \\
  & & mean & $1472.852\pm0.006$ & & 0.34\\
  \noalign{\smallskip}
  \hline
\end{tabular}

\flushleft{\small $^a$Difference with model frequency assumes the
$m=-1$ component lies $\pm0.4\,\rm \mu Hz$ from central frequency.\\
$^b$Difference with model frequency assumes $m=-2$ components lies
$\pm0.8\,\rm \mu Hz$ from central frequency.}
\end{table}

\subsection{Results of searching smoothed spectra with Bayesian
statistics}\label{section[results bayes smooth]} As previously
mentioned it is likely that the power of the p~modes is spread over
more than one frequency bin. Furthermore, \citet{Garcia2007} found
evidence that could suggest that g~modes have short lifetimes, and
so g-mode powers could also be spread across more than one frequency
bin. With this in mind we have used Bayesian techniques to search
smoothed frequency-power spectra.

\begin{figure}
\centering
  \includegraphics[width=\columnwidth, clip]{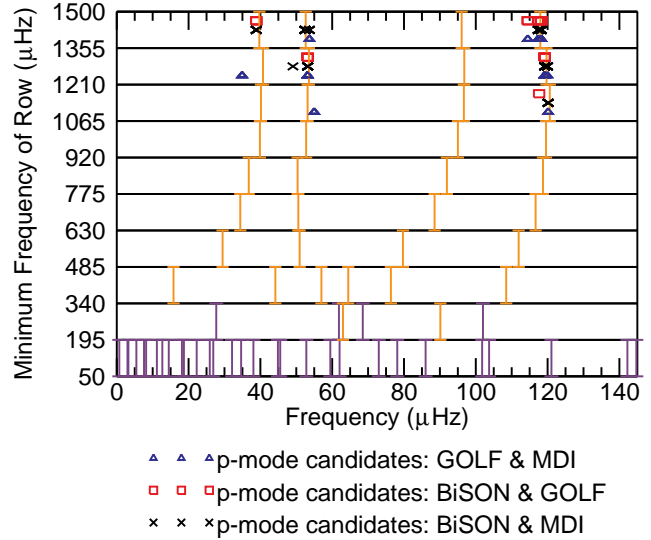}\\
  \caption{An echelle plot, modulo $145\,\rm \mu Hz$, marking locations in
frequency of
  occurrences uncovered by the Bayesian test searches of smoothed spectra. The
frequency-power spectra were
  smoothed over two, four, six, eight, 10, 12, 14, 16, 18 and 20 bins. The different
  symbols represent the different pairings of data that were searched (see
legend).}
  \label{figure[echelle smoothed]}
\end{figure}

\begin{table*}
\caption{Candidates found when using Bayesian statistics to search
smoothed spectra. $R$ is the number of bins we have smoothed
over.}\label{table[smoothed bayesian results]}
\begin{tabular}{ccccccc}
  \hline
  \noalign{\smallskip}
  \small{\emph{l}} &\small{\emph{n}} & \small{\emph{m}} & \small{Frequency} &
   \small{Minimum} & \small{$R$ at which} & Distance \\
   &  &  & \small{($\mu$Hz)} & \small{Posterior}
   & \small{minimum} & from model \\
   &  &  &  & \small{probability} &  $P(\textrm{H}00|x)$ & frequency  \\
   &  &  &  & \small{$[P(\textrm{H}00|x)]$} & obtained  & ($\mu$Hz) \\
   \noalign{\smallskip}
  \hline
  \noalign{\smallskip}
  0 & 7 & 0 & $1120.01\pm0.05$ & $4.4\times10^{-3}$ & 14 & 1.62 \vspace{2mm}\\
  1 & 7 & $-1$ & $1182.64\pm0.02$ & $5.4\times10^{-3}$ & 12 & 2.82$^a$ \\
  1 & 7 & $-1$ & $1185.17\pm0.02$ & $1.2\times10^{-3}$ & 12 & 0.29$^a$ \vspace{2mm}\\
  2 & 7 & $-2$ & $1244.90\pm0.06$ & $1.3\times10^{-3}$ & 16 & 5.07$^b$ \vspace{2mm}\\
  0 & 8 & 0 & $1259.08\pm0.05$ & $7.2\times10^{-3}$ & 14 & 4.52 \\
  1 & 8 & $-1$ & $1329.17\pm0.08$ & $6.4\times 10^{-25}$ & 20 & \\
  1 & 8 & $+1$ & $1329.93\pm0.08$ & $4.2\times10^{-4}$ & 20 & \\
  & & mean & $1329.55\pm0.11$ & & & 0.36 \vspace{2mm}\\
  2 & 8 & $-2$ & $1393.87\pm0.07$ & $1.3\times10^{-4}$ & 18 & 0.34$^b$ \vspace{2mm}\\
  0 & 9 & 0 & $1407.36\pm0.08$ & $2.4\times10^{-12}$ & 18 & 0.63$^b$ \vspace{2mm}\\
  1 & 9 & $-1$ & $1472.43\pm0.04$ & $8.2\times 10^{-18}$ & 10 & \\
  1 & 9 & $+1$ & $1473.19\pm0.07$ & $2.5\times 10^{-12}$ & 18 & \\
  & & mean & $1472.81\pm0.08$ & & & 0.38 \\
  \noalign{\smallskip}
  \hline
\end{tabular}

\flushleft{\small $^a$Difference with model frequency assumes the
$m=-1$ component lies $\pm0.4\,\rm \mu Hz$ from central frequency.\\
$^b$Difference with model frequency assumes $m=-2$ components lies
$\pm0.8\,\rm \mu Hz$ from central frequency.}
\end{table*}

We have used the methods described in Section \ref{section[smoothed
bayes]} to search smoothed frequency-power spectra. We have then
used equation (\ref{equation[posterior 2 spectra]}) to determine the
posterior probability. Again the frequency-power spectra were
searched in pairs with one time series shifted to start $24\,\rm h$
after the other.

Fig. \ref{figure[echelle smoothed]} shows the results of searching
the BiSON, GOLF and MDI data when the frequency-power spectra were
smoothed over various numbers of bins (all even values between 2 and
20). The uneven prior has been applied and removes 17 false
detections, some of which can be attributed to daily harmonics and
spacecraft frequencies.

Table \ref{table[smoothed bayesian results]} contains the
frequencies of the observed mode candidates. Notice that the number
of bins we needed to smooth over, $R$, to minimize the observed
posterior probability, $P(\textrm{H}00|x)$, is reasonably large
(column 5). This highlights the truly resolved nature of the p~mode
profiles in the frequency-power spectra. Theoretically, the minimum
value of $p(\textrm{H}00|x)$ will occur when $R$ is equal to the
linewidth of the mode \citep{Appourchaux2009}. As can be seen in
Table \ref{table[smoothed bayesian results]} $R$ does, in general,
decrease with mode frequency. However, there are quite a few cases
where this is not the case. This could be because we are looking for
modes whose powers are only marginally greater than the background
noise. Consequently, the mode profiles are difficult to distinguish
from the noise, particularly in the limbs of the profiles and so the
widths of the modes remain uncertain. The errors on the frequencies,
given in Table \ref{table[smoothed bayesian results]}, are the bin
width of the smoothed spectra in which the minimum probabilities
were observed (i.e. $0.004R\,\rm\mu Hz$). These error estimates are
significantly larger than those given in Table \ref{table[Bayesian
results]}.

Although formally allowed by the uneven prior, we regard any
detections that are further than $1\,\rm\mu Hz$ from the predicted
frequencies with suspicion as, in general, the difference is
significantly less than this. Notice that there are two $l=1$,
$n=7$, $m=-1$ candidates and two $l=0$, $n=8$ candidates. In each
case the candidate that lies closest to the predicted mode frequency
also has the lowest posterior probability and so these candidates
are more likely to represent the actual modes. No g-modes candidates
were detected regardless of the assumed width and power. Once again
we note that the results described in this section are dependent on
the initial assumptions made when using the Bayesian approach.

\section{Discussion of the results}\label{section[discussion]}
We begin by comparing the results observed in the unsmoothed spectra
using frequentist and Bayesian search methods. The frequencies of
candidates that are detected by both approaches show very good
agreement with each other and with previously claimed detections
\citep[see e.g.][]{Toutain1998, Bertello2000, Garcia2001,
Chaplin2002, Broomhall2007, Salabert2009}. When the frequentist
approach was applied, more false detections (i.e. detections that
were found to lie away from the predicted mode frequencies) were
observed than expected. We believe this is significant and
highlights the misleading nature of the frequentist approach. When
the Bayesian approach was applied significantly fewer false
detections were observed, even before the uneven prior probability
was applied. Furthermore, the Bayesian approach allows us to use our
knowledge of the Sun in a statistically rigourous manner to guide
the frequencies at which we consider detections to be mode
candidates (with the uneven prior probability). Although Bayesian
methods may lead to more stringent threshold levels and consequently
fewer detections we feel that this is a small price to pay for the
improved accuracy and confidence one achieves in any detections. It
is possible that using the uneven prior probability may discount
detections that are in fact real mode signatures (type~II errors).
However, we feel that use of the uneven prior probability is
justified as it also helps to minimize the number of noise
signatures that are considered as mode candidates (type~I errors).
The Bayesian approach is also less open to misinterpretation than
the frequentist approach. However, once again we note that the
Bayesian approach is reliant on the assumptions made when
determining the posterior probability. Different assumptions may
lead to different mode candidates being uncovered. For example, the
assumption that the rotational splitting of g~modes is $0.4\,\mu Hz$
may be incorrect \citep{Garcia2007}.

Mode candidates were also detected in smoothed spectra. This method
was particularly successful at higher frequencies, where the modes
have broader widths. It is worth noting that as the p-mode powers
and widths were determined by extrapolation it is likely that the
values used here are more accurate higher in the frequency search
range. This may have influenced the results especially if the mode
widths were overestimated. We were able to eliminate mode candidates
that were positioned away from the predicted mode frequencies by
employing the non-uniform prior probability. This cannot be done for
the frequentist approach as we are effectively using a priori
knowledge to guide the regions in frequency we search. The use of a
priori knowledge in a statistically rigourous manner can only be
done through Bayesian statistics.

The g-mode candidate that was observed in the frequentist approach
was not detected when Bayesian statistics were used. One g~mode
candidate was detected using the Bayesian approach. However, each
detection was only made in one combination of the data sets and
neither candidate was detected with both the frequentist and
Bayesian approaches. Therefore, we regard these two candidates with
suspicion and feel that each candidate still requires further
investigation and independent confirmation. The candidate that was
detected with the Bayesian approach was not detected when the
spectra were smoothed. This could indicate that, if it is a g~mode,
it has a lifetime that is longer than the lifetimes assumed here.

\section*{Acknowledgements}

The authors thank S. Basu and A. Serenelli for the solar models that
were used to produce Fig. \ref{figure[model echelle]}. We would like
to thank all members of the Phoebus collaboration for much valued
discussions concerning this work. This paper utilizes data collected
by the Birmingham Solar-Oscillations Network (BiSON). We thank the
members of the BiSON team, both past and present, for their
technical and analytical support. We also thank P. Whitelock and P.
Fourie at SAAO, the Carnegie Institution of Washington, the
Australia Telescope National Facility (CSIRO), E.J. Rhodes (Mt.
Wilson, Californa) and members (past and present) of the IAC,
Tenderize. BiSON is funded by the Science and Technology Facilities
Council (STD). The authors also acknowledge the financial support of
STD. We thank the referee for insightful comments.
\small
\bibliographystyle{mn2e}
\bibliography{bayes_paper_v4}

\appendix
\begin{table*}
\flushleft\section{Table of assumptions}\label{section[appendix]}
\centering \caption{Assumptions made with each
approach.}\label{table[assumptions]}
\begin{tabular}{ccc}
  \hline
  \noalign{\smallskip}
 Property & Frequentist & Bayesian \\
   \noalign{\smallskip}
  \hline
     \noalign{\smallskip}
  Underlying statistics & Gaussian  & $\chi^2$ 2d.o.f. \\
   & (for frequency-amplitude spectra) & (for frequency-power
   spectra)\\
      & & \\
  Threshold probability & 0.01 & 0.01 \\
     & & \\
  Rotational splitting & $0.4\,\rm\mu Hz$ & $0.4\,\rm\mu Hz$\\
     & & \\
  Power of p~modes & None & Based on extrapolations\\
  & & (see Section \ref{section[amplitude range]}):\\
      & & $0.1\,\textrm{mm}^2\,\textrm{s}^{-2}\leq V^2 \leq 38\,\textrm{mm}^2\,\textrm{s}^{-2}$\\
      & & \\
  Power of g~modes & None & Based on models \\
  & & (see Section \ref{section[amplitude range]}):\\
      & & $V^2=9$ or $36\,\textrm{mm}^2\,\textrm{s}^{-2}$\\
      & & \\
  Frequencies considered  & Closer than $1\,\rm\mu Hz$  & Allowed by uneven prior probability \\
  as p-mode candidates & to a model mode frequency & i.e. an inverted top-hat distribution, \\
     & & based on frequencies predicted by \\
     & & 5000 models \\
     & & \\
  Frequencies considered  & Closer than $1\,\rm\mu Hz$  & Allowed by uneven prior probability \\
  as g-mode candidates & to a model mode frequency & i.e. an inverted top-hat distribution, \\
     & & based on frequencies predicted by \\
     & & models and an uncertainty of 1 per cent\\
     & & \\
  Lifetimes of p~modes & Long enough for the power to  & Based on
  extrapolations\\
    & be confined to a few bins only & (see Section \ref{section[amplitude range]}) \\
   & & $795\,\textrm{d}\geq\tau\geq35\,\textrm{d}$\\
      & & \\
  Lifetimes when searching  & Greater than the length of & Greater than the length of \\
  for g-mode single spikes & the time series i.e. $\tau>3071\,\rm d$. & the time series i.e. $\tau>3071\,\rm d$ \\
     & & \\
  Lifetimes when searching & 1, 2, 3, and 4 months & 1, 2, 3, and 4 months \\
  for g-mode peaks & & \\
     & & \\
  \hline
\end{tabular}
\end{table*}

\end{document}